\documentclass[aps,floats]{revtex4}
\usepackage{amsmath,amssymb}
\usepackage{graphicx,epsfig}

\begin{document}
\bibliographystyle {plain}

\def\oppropto{\mathop{\propto}} 
\def\opsimeq{\mathop{\simeq}}
\def\opoverderline{\mathop{\overline}}
\def\operarrow{\mathop{\longrightarrow}}
\def\opsim{\mathop{\sim}}

\def\fig#1#2{\includegraphics[height=#1]{#2}}
\def\figx#1#2{\includegraphics[width=#1]{#2}}


\title{ Directed polymer in a random medium of dimension $1+1$ and $1+3$: \\
 weights statistics in the low-temperature phase   } 


\author{ C\'ecile Monthus and Thomas Garel }
 \affiliation{Service de Physique Th\'{e}orique, CEA/DSM/SPhT\\
Unit\'e de recherche associ\'ee au CNRS\\
91191 Gif-sur-Yvette cedex, France}

\begin{abstract}
We consider the low-temperature $T<T_c$ disorder-dominated phase
of the directed polymer in a random potentiel in dimension
$1+1$ (where $T_c=\infty$) and $1+3$ (where $T_c<\infty$). 
To characterize the localization properties of the polymer of length $L$,
 we analyse the statistics of the 
 weights $w_L(\vec r)$ of the last monomer as follows.
We numerically compute the probability
distributions $P_1(w)$ of the maximal weight $w_L^{max}= max_{\vec r}
[w_L(\vec r)]$, the probability distribution $\Pi(Y_2)$ of the parameter
$Y_2(L)= \sum_{\vec r} w_L^2(\vec r) $
as well as the average values of the higher order
moments $Y_k(L)= \sum_{\vec r}  w_L^k(\vec r)$.
We find that there exists a temperature $T_{gap}<T_c$ such that
(i) for $T<T_{gap}$, the distributions $P_1(w)$ and
$\Pi(Y_2)$ present the characteristic Derrida-Flyvbjerg singularities
at $w=1/n$ and $Y_2=1/n$ for $n=1,2..$. In particular, there exists a
temperature-dependent exponent $\mu(T)$ that governs the main singularities
$P_1(w) \sim (1-w)^{\mu(T)-1}$ and $\Pi(Y_2) \sim (1-Y_2)^{\mu(T)-1}$
as well as the power-law decay of the  moments $ \overline{Y_k(i)}
\sim 1/k^{\mu(T)}$.  The exponent $\mu(T)$ grows from the value
$\mu(T=0)=0$ up to $\mu(T_{gap}) \sim 2$.
(ii) for $T_{gap}<T<T_c$, the
distribution $P_1(w)$ vanishes at some value $w_0(T)<1$, and
accordingly the moments $\overline{Y_k(i)}$ decay exponentially as
$(w_0(T))^k$ in $k$.  The histograms of spatial correlations also display 
 Derrida-Flyvbjerg singularities for $T<T_{gap}$.
Both below and above $T_{gap}$, the study of typical and averaged
correlations is in full agreement with the droplet scaling theory.

\end{abstract}

\maketitle

\section{Introduction}

A convenient way to characterize disorder-dominated phases
is through the statistics of some appropriate ``weights''. 
In mean-field models, these weights represent either
weights of pure states, as in the replica analysis
of the Sherrington-Kirkpatrick model \cite{replica},
or weights of microscopic configurations, as in the 
Random Energy Model \cite{rem} or in the directed polymer
model on the Cayley tree \cite{Der_Spo}.
It turns out that in these three cases, the weights statistics
is the same as in L\'evy sums with some index $0<\mu<1$  \cite{Der},
where the index $\mu$ depends on the temperature :
for instance in the Random Energy Model \cite{rem}
 or in the directed polymer model on the Cayley tree \cite{Der_Spo},
it is simply $\mu(T)=T/T_c$.
In \cite{Der_Fly}, the corresponding probability distributions
of the weights were found to exhibit characteristic singularities
at some integer inverses.
Similar Derrida-Flyvbjerg singularities 
also occur in many other contexts, such as
randomly broken objects \cite{Der_Fly,Kra},
in population genetics \cite{Higgsbio,Der_Jun,Dertrieste},
in random walk excursions or loops \cite{Der,Fra,Kan}.

For disordered systems in finite dimensions, 
it seems appropriate to consider the weights associated
to a local degree of freedom, to characterize to what extent it is
frozen. To the best of our knowledge, 
this idea has first been introduced 
to characterize the freezing of a folded polymer
with random self-interactions \cite{De_Gr_Hi}.
It was then used in the context of secondary structures
of random RNA to analyse the fraction of frozen
pairs between degenerate ground states \cite{Mor_Hig},
and to characterize the freezing transition \cite{rnapoids}.

In this paper, we study the statistics of the 
 weights $w_L(\vec r)$ of the end-point of a directed 
polymer in a random potentiel \cite{Hal_Zha}.
We focus here on the 
low-temperature $T<T_c$ disorder-dominated phase
both in dimension
$1+1$ (where $T_c=\infty$) and $1+3$ (where $T_c<\infty$),
since we have studied elsewhere \cite{DPmultif}
the weights statistics
at criticality in $d=3$, where multifractal behavior occurs.
We are not aware of previous studies on these weights in
the physics literature. On the contrary,
in the mathematical litterature, the weight of the favourite site
has been considered as a localization criterion \cite{Car_Hu,Com}, and
a more detailed description of end-point weights
was then given via the notion of $\epsilon$-atoms \cite{Var}.

The paper is organized as follows.
The model and observables are introduced in Section \ref{model}.
We then present a detailed study of the weights of the end-point
of the directed polymer both in dimensions $1+1$
and $1+3$. For clarity, the statistical properties of the weights
alone, independently of the distances involved are described
in Section \ref{poids}, whereas the study of spatial properties
is postponed to Section \ref{spatial}. We summarize our results in Section
\ref{conclusion}.

\section{ Model and observables }

\label{model}

\subsection{ Model definition} 

In this paper, we present numerical results
for the random bond version
of the model defined by the recursion relation
on a cubic lattice in $d=1$ and $d=3$
\begin{eqnarray}
\label{DP1}
Z_{t+1} (\vec r) =  \sum_{j=1}^{2d}
 e^{-\beta \epsilon_t(\vec r+\vec e_j,\vec r)} Z_{t} (\vec r+\vec e_j)
\label{transfer}
\end{eqnarray}
The bond energies $\epsilon_t(\vec r+\vec e_j,\vec r) $
are random independent variables drawn from the Gaussian
distribution 
\begin{eqnarray}
\rho (\epsilon) = \frac{1}{\sqrt{2\pi} } e^{- \frac{\epsilon^2}{2} }
\end{eqnarray}
In this paper, we consider the following boundary conditions.
 The first monomer is fixed at $\vec r = \vec 0$, 
i.e. the initial condition of the recurrence of Eq. (\ref{DP1}) reads
\begin{eqnarray}
Z_{t=0} (\vec r) = \delta_{\vec r, \vec 0}
\end{eqnarray}
The last monomer is free, i.e. the full partition function of the polymer of length $L$  
is then obtained by summing over all possible positions $\vec r$
at $t=L$
\begin{eqnarray}
Z_L^{tot} = \sum_{\vec r} Z_L(\vec r)  
\label{ztot}
\end{eqnarray}

This model has attracted a lot of attention because it is directly related
to non-equilibrium properties of growth models 
\cite{Hal_Zha}.
Within the field of disordered systems, it is also very interesting on its own
because it represents a `baby-spin-glass' model
\cite{Hal_Zha,Der_Spo,Der,Mez,Fis_Hus}.  At low
temperature, there exists a disorder dominated phase, where the 
order parameter is an `overlap'.
In finite dimensions, a scaling droplet theory was proposed
 \cite{Fis_Hus,Hwa_Fis},
in direct correspondence with the droplet
 theory of spin-glasses \cite{Fis_Hus_SG},
whereas in the mean-field version of the model on the Cayley,
a freezing transition very similar to the one occurring
in the Random Energy Model was found \cite{Der_Spo}.
The phase diagram as a function of space dimension $d$ is the
following \cite{Hal_Zha}. In dimension $d \leq 2$, there is no free phase,
i.e. any initial disorder drives the polymer into the strong disorder phase,
whereas for $d>2$, 
there exists a phase transition between
the low temperature disorder dominated phase
and a free phase at high temperature  \cite{Imb_Spe,Coo_Der}.

In the following, we  will focus on the statistical properties 
of the weights   
\begin{eqnarray}
w_L (\vec r) = \frac{ Z_L(\vec r) }{ Z_L^{tot}} 
\label{defw}
\end{eqnarray}
normalized to (Eq. \ref{ztot})
\begin{eqnarray}
\sum_{\vec r} w_L (\vec r) = 1
\label{norma}
\end{eqnarray}

The numerical results given below have been obtained using polymers 
of various lengths $L$,
with corresponding numbers $n_s(L)$ of disordered samples with
the values
\begin{eqnarray}
L && = 50,100,200,400, 800 \\
n_s(L) && = 13.10^7 , 35.10^6 , 9.10^6 , 225.10^4 , 57.10^4
\label{nume1d}
\end{eqnarray}
in $d=1$, and the values
\begin{eqnarray}
L && =6, 12,18,24,36,48,60 \\
n_s(L) && = 10^8, 10^7, 2.10^6, 8.10^5, 2.10^5, 5. 10^4, 3. 10^4
\label{nume3d}
\end{eqnarray}
in $d=3$. In the following, $\overline{A}$ denotes the average of $A$
over the disorder samples.

\subsection{Characterization of the weights statistics}

In analogy with the weight statistics in L\'evy sums and in
the Random Energy Model \cite{Der,Der_Fly},
we have numerically computed the probability distribution 
$P_1(w)$ of the maximal weight (Eq. \ref{defw})
\begin{equation}
w_L^{max}= max_{\vec r} \{ w_L( \vec r) \}
\label{wmax}
\end{equation}
as well as  the probability distribution $P_2(w)$ of the second maximal weight.
Another useful way to characterize the statistical
properties of the weights  \cite{Der,Der_Fly}
is to consider the moments of arbitrary order $k$
\begin{equation}
Y_k(L)  = \sum_{ \vec r} w_L^k(\vec r)
\label{ykdef}
\end{equation}
which represents the probability that the last monomer of the polymer
of length $L$
is at the same point in $k$ different thermal configurations
of the same disordered sample.
We have measured the probability $\Pi(Y_2)$ of the parameter
\begin{equation}
Y_2(L)= \sum_{ \vec r} w_L^2(\vec r)
\label{y2def}
\end{equation}
 as well as the moments $\overline{Y_k(L)}$ for $2 \leq  k \leq 100$.
Finally, we have also computed 
the weights density
\begin{equation}
f_L(w) = \overline{ \sum_{ \vec r} \delta(w-w_L(\vec r)) }
\label{densityfdef}
\end{equation}
giving rise to the moments
\begin{equation}
\overline{Y_k(L)} = \int_0^1 dw w^k f_L(w)
\label{ykf}
\end{equation}
The normalization condition for the density $f_L(w)$ is
\begin{equation}
\overline{Y_1(L)} = \int_0^1 dw w f_L(w) =1
\label{normay1}
\end{equation}

In the following, we will also present histograms of
the associated entropy
\begin{equation}
s_L= - \sum_{ \vec r } w_L(\vec r) \ln  w_L(\vec r)
\label{entropy}
\end{equation}

\section{ Study of the weights statistics} 

\label{poids}

\subsection{Probability distribution $P_1(w)$
 of the largest weight }

\begin{figure}[htbp]
\includegraphics[height=6cm]{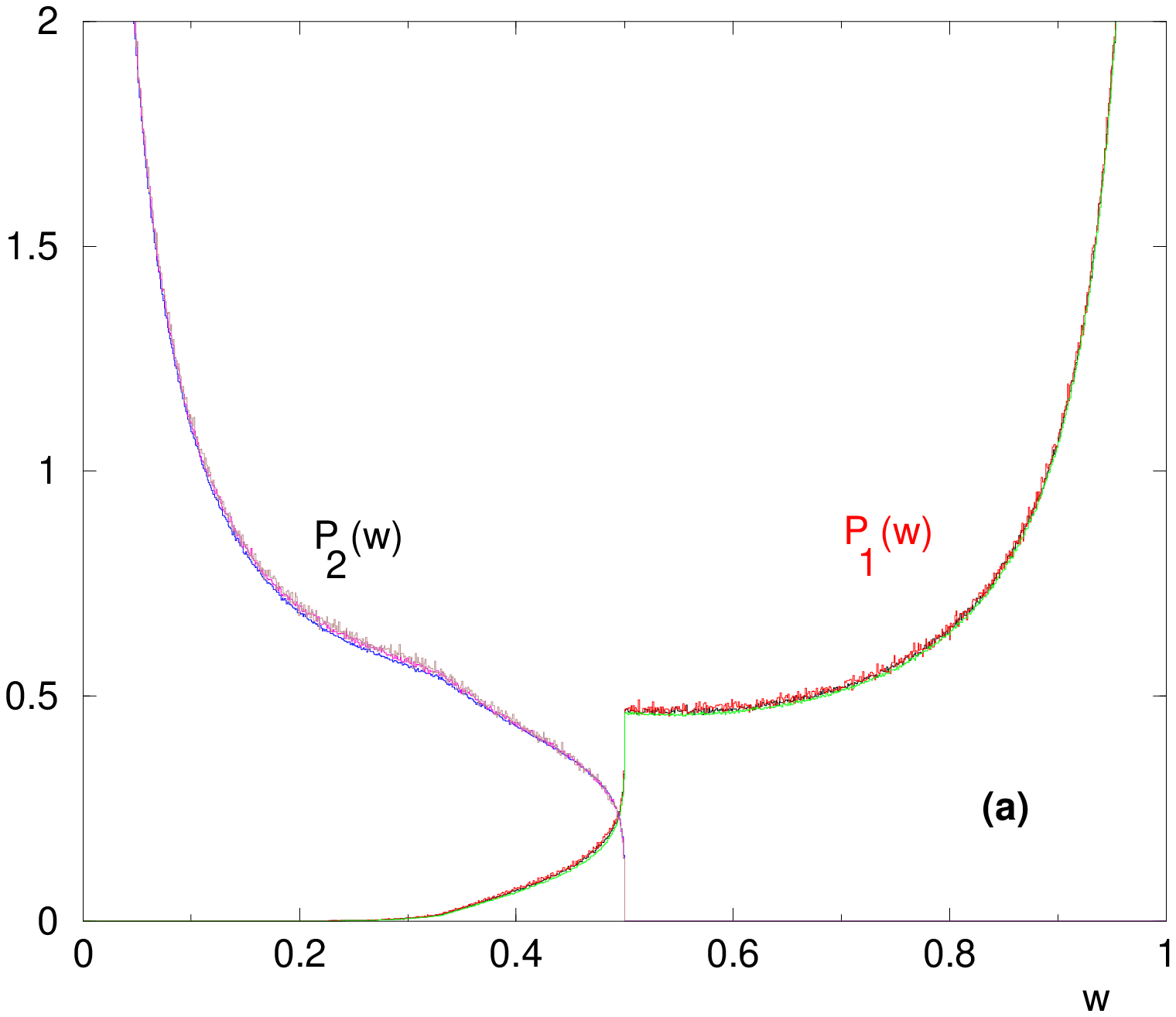}
\hspace{1cm}
\includegraphics[height=6cm]{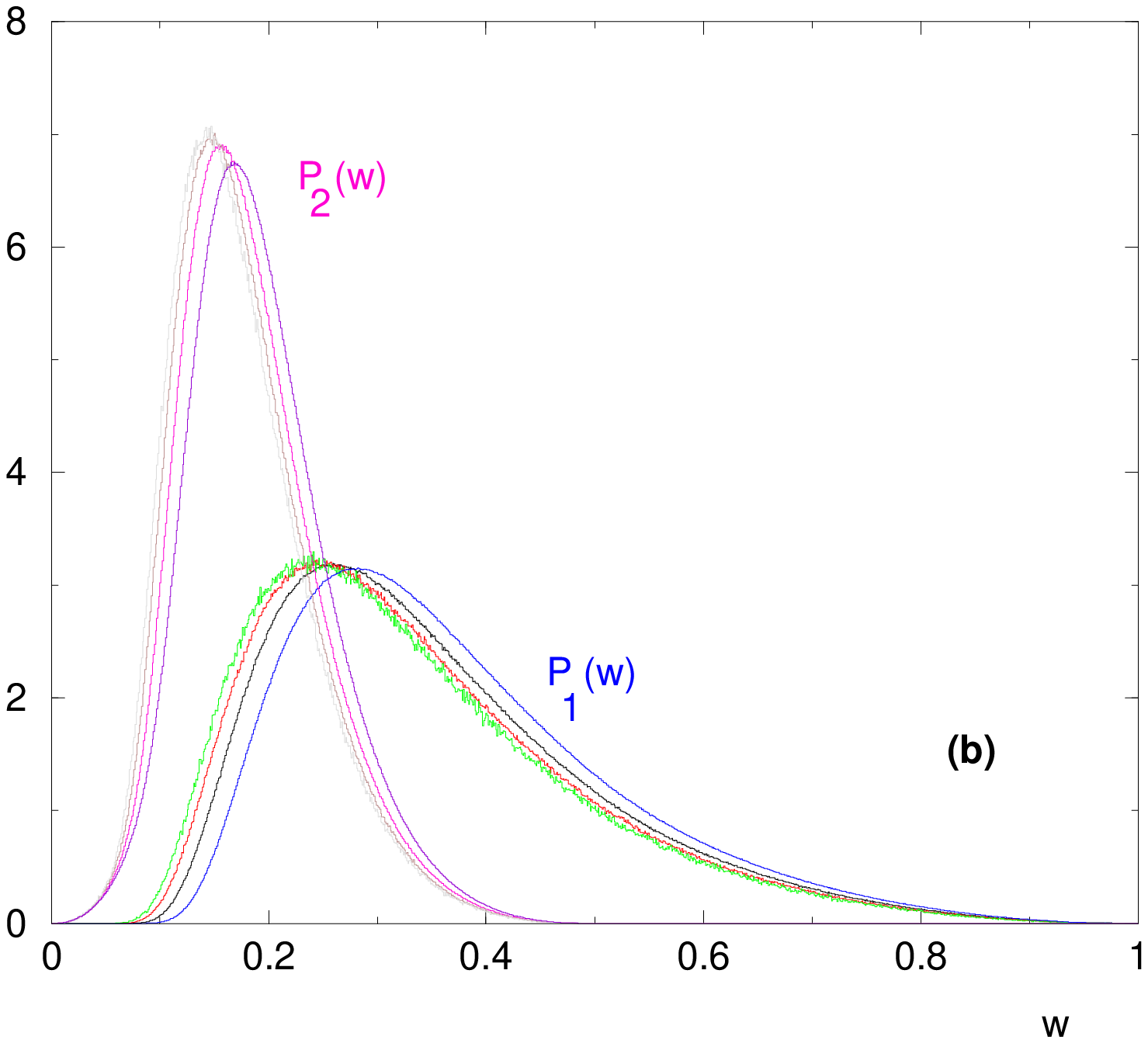}
\caption{(Color online) $d=1$ : 
 Probability distributions $P_1(w)$ and $P_2(w)$
 of the largest and second largest
 weight seen by the last monomer (see Eq \ref{wmax})
(a) at $T=0.1$ ($T<T_{gap}$) for $L=50,100,200$ :
the characteristic Derrida-Flyvbjerg 
singularities at $w=1$ and $w=1/2$ are clearly visible.
 (b) at $T=1.$ ($T>T_{gap}$) for $L=50,100,200,400$ :
the distribution $P_1(w)$ does not reach $w=1$ anymore, 
and the distribution $P_2(w)$ does not reach $w=1/2$ anymore  }
\label{fig1dw}
\end{figure}

\begin{figure}[htbp]
\includegraphics[height=6cm]{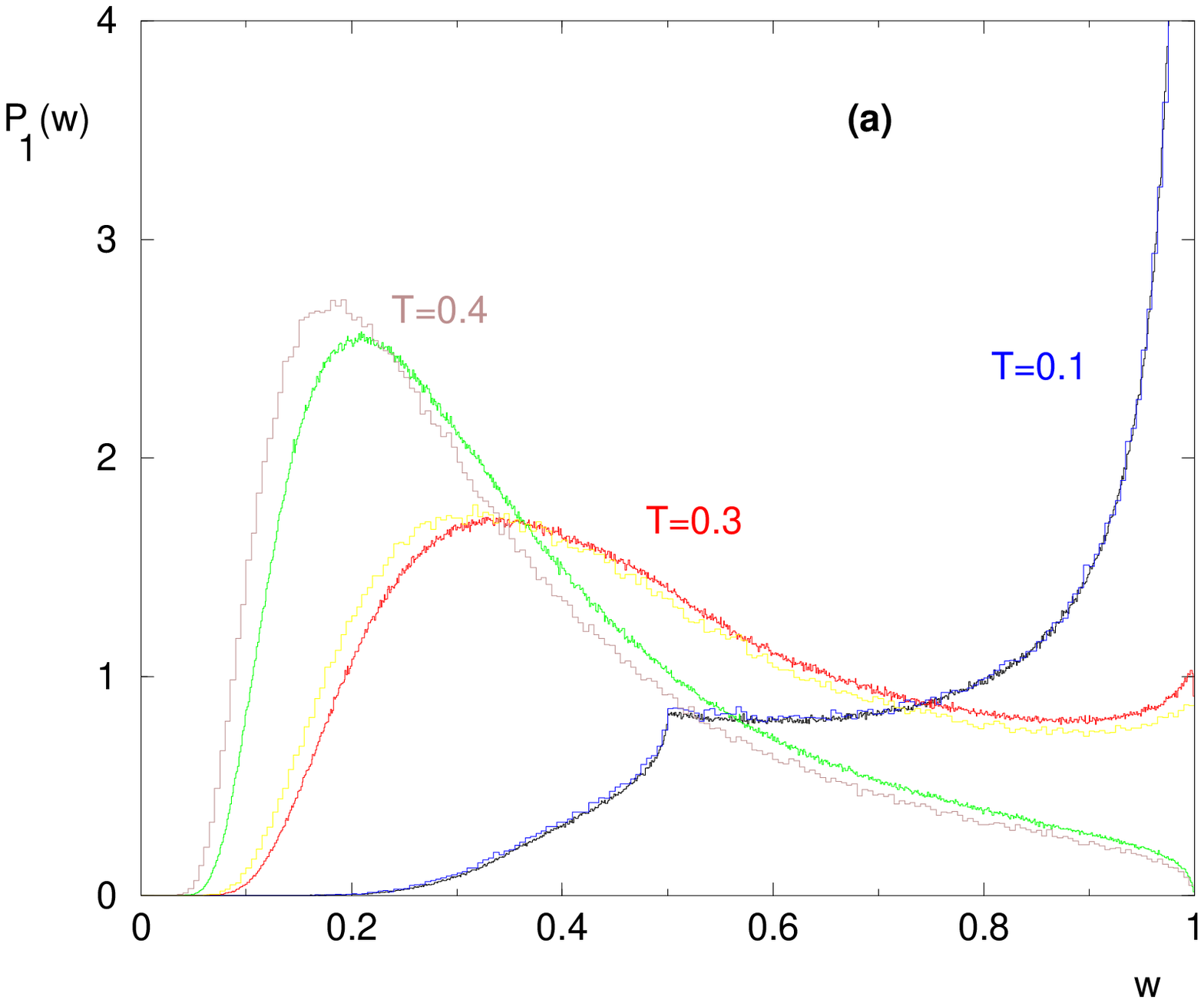}
\hspace{1cm}
\includegraphics[height=6cm]{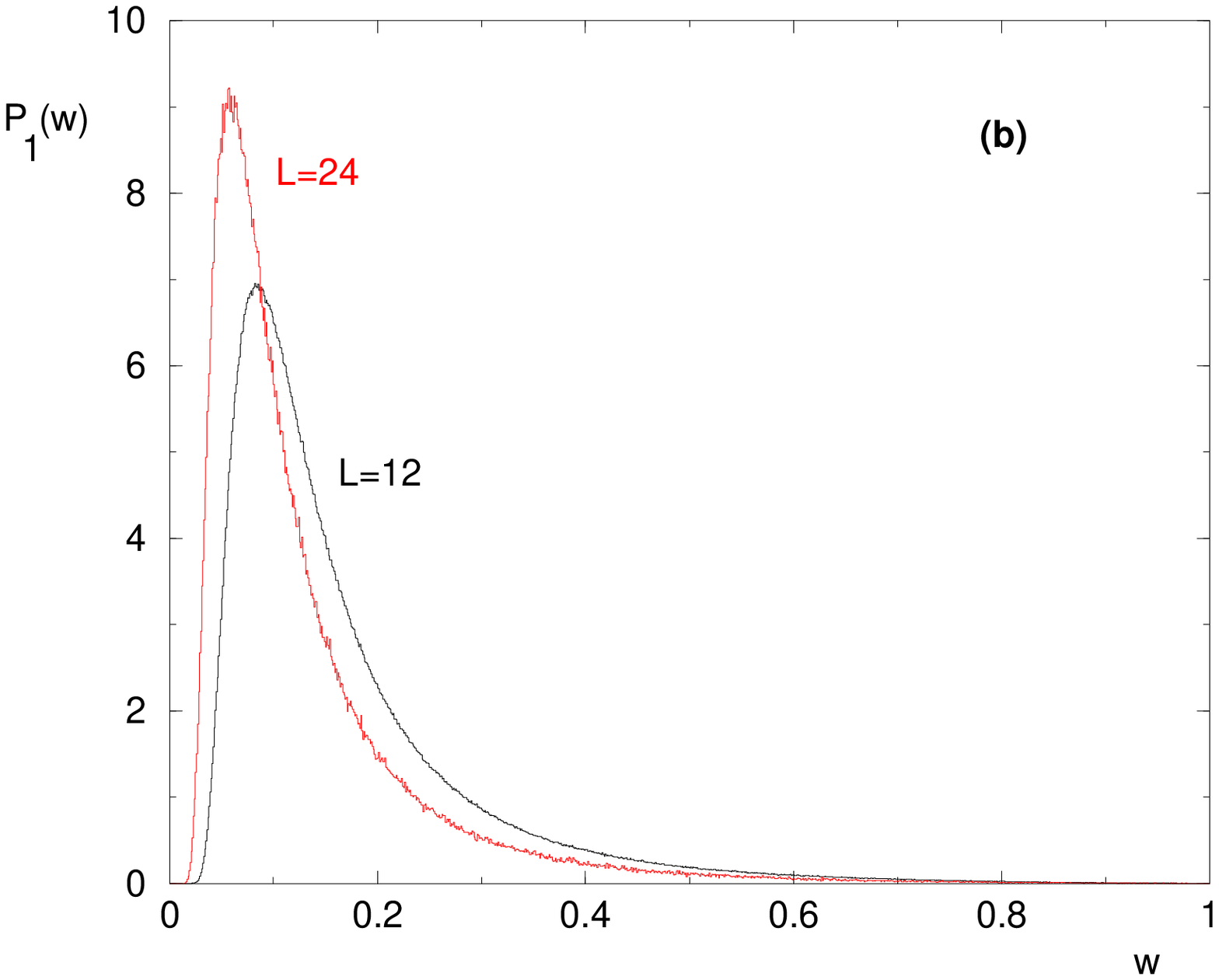}
\caption{(Color online) $d=3$ :
 Probability distribution $P_1(w)$
 of the largest weight seen by the last monomer (see Eq \ref{wmax})
(a) at $T = 0.1$ ($\mu<1$) , $T=0.3$ ($\mu \sim 1$) ,
$T=0.4$ ($1<\mu<2$) for $L=12,24$.
 (b) at $T=0.6 > T_{gap}$  for $L=12,24$ 
 }
\label{fig3dw}
\end{figure}

The probability distributions $P_1(w)$ and $P_2(w)$ of the largest
(Eq. \ref{wmax})
and second largest weights of the last monomer
in dimension $1+1$ are shown on Fig. \ref{fig1dw} for two temperatures.
These curves show that there exists a temperature $T_{gap}(d=1) \sim 0.7$ such that

(i) for $T<T_{gap}$ (see Fig. \ref{fig1dw} a )  the distribution $P_1(w)$
reaches the point $w \to 1$ with a singularity parametrized
by a temperature-dependent exponent 
\begin{equation}
P_1(w) \oppropto_{w \to 1} (1-w)^{\mu(T)-1}
\label{p1singw1}
\end{equation}
Beyond this main singularity, $P_1(w)$ also present characteristic 
 Derrida-Flyvbjerg singularities at $w=1/2,1/3...1/n...$
\cite{Der_Fly} : in particular, the singularity of $P_1(w)$ at $w=1/2$
is clealy visible on Fig.  \ref{fig1dw} a.
Similarly, the distribution $P_2(w)$
reaches the point $w \to 1/2$ (see Fig. \ref{fig1dw} a )

(ii)  for $ T> T_{gap}  $  (see Fig. \ref{fig1dw} b )
the distribution $P_1(w)$ does not reach $w=1$ anymore, but vanishes
at some maximal value $0<w_0(T)<1$  
\begin{equation}
P_1(w) \oppropto_{w \to w_0(T)} (w_0(T)-w)^{\sigma}
\label{p1gapw0}
\end{equation}
with some exponent $\sigma$. Similarly, the distribution $P_2(w)$
does not reach the point $w = 1/2$ (see Fig. \ref{fig1dw} b )

This temperature $T_{gap}$ in $1+1$ where $T_c=\infty$ 
also exists in $1+3$ where $T_c $ is finite,
as shown on Fig. \ref{fig3dw}.
On Fig. \ref{fig3dw} a, the distribution $P_1(w)$
is shown for three temperatures below $T_{gap}$
with exponents  $\mu(T=0.1)<1$ , $\mu(T=0.3) \sim 1$
and $1<\mu(T=0.4)<2$. On Fig. \ref{fig3dw} b, 
the distribution $P_1(w)$ is shown for temperature
$T \sim 0.6$ in the region $T_{gap}(d=3) \sim 0.5 <T < T_c \sim 0.79 $.

Note that  here we only describe the low-temperature
phase $T<T_c$ and we refer to \cite{DPmultif}
for a detailed study of the weights statistics
at criticality $T_c=0.79$ where multifractal behavior occurs.

\subsection{ Probability distribution $G_L(s)$ of the entropy
$s_L= - \sum_{ \vec r } w_L(\vec r) \ln  w_L(\vec r)$   }

\begin{figure}[htbp]
\includegraphics[height=6cm]{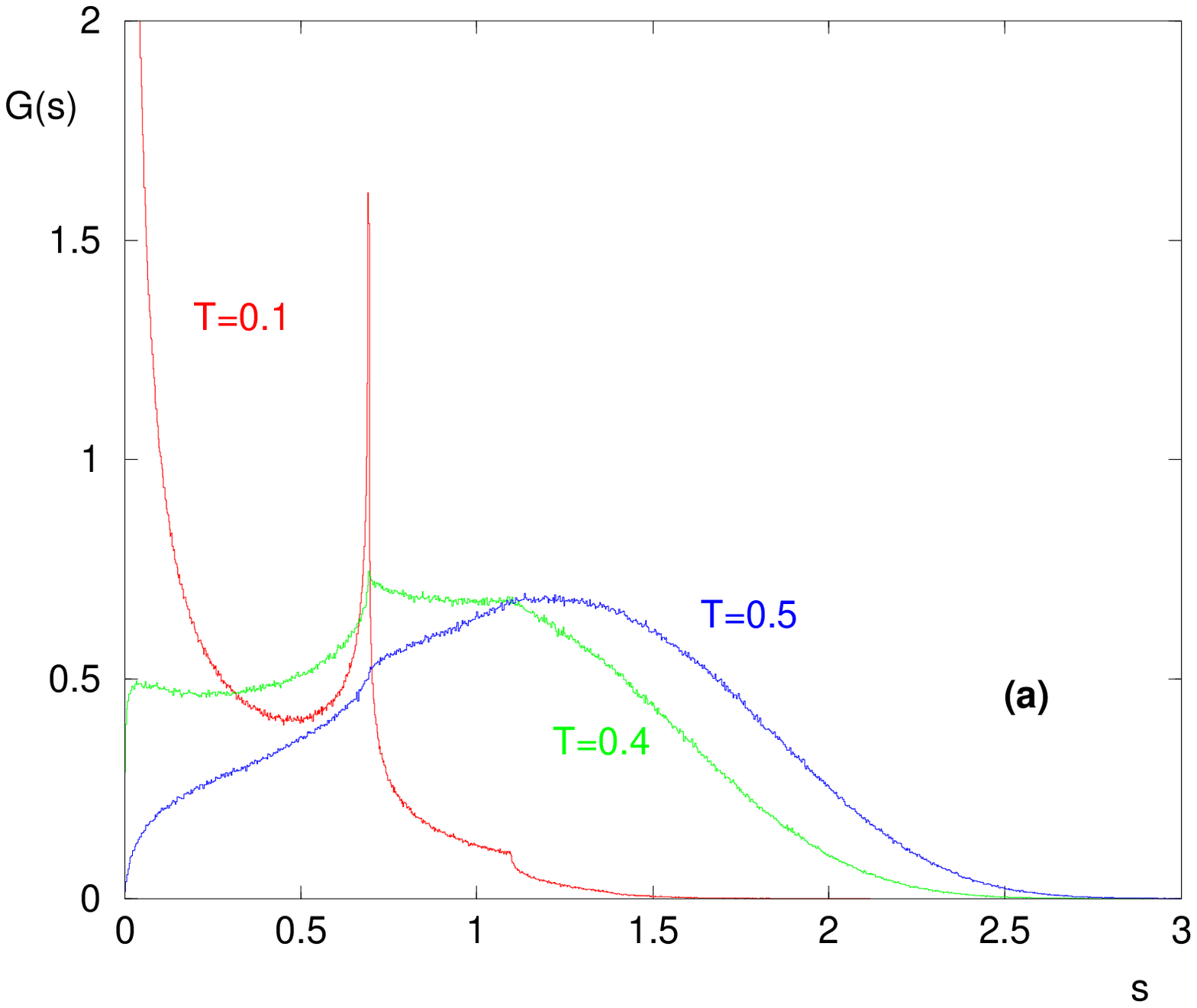}
\hspace{1cm}
\includegraphics[height=6cm]{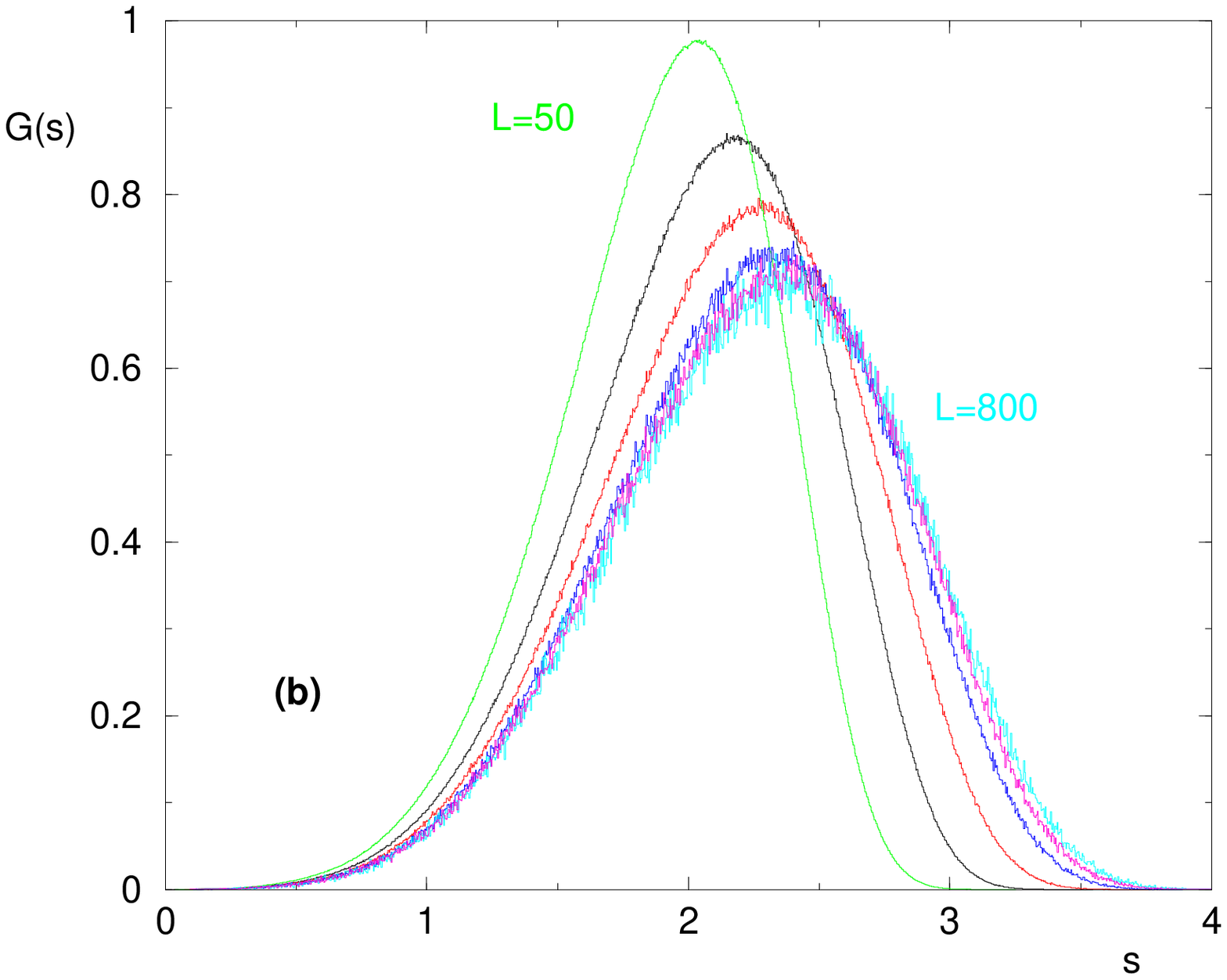}
\caption{(Color online) $d=1$ : 
Probability distribution $G_L(s)$ of the last monomer entropy
$s_L= - \sum_{ \vec r } w_L(\vec r) \ln  w_L(\vec r)$
(a) at three temperatures below $T_{gap}$ namely
$T = 0.1$ ($\mu<1$) , $T=0.4$ ($\mu \sim 1$) ,
$T=0.5$ ($1<\mu<2$) for $L=200$ :
the Derrida-Flyvbjerg 
singularities at $s=0$, $s=\ln 2$ and $s=\ln 3$ are clearly visible.
 (b) at $T=1. >T_{gap}$ for $L=50,100,200,400,600,800$ : the histogram
does not reach $s=0$. }
\label{fig1ds}
\end{figure}

\begin{figure}[htbp]
\includegraphics[height=6cm]{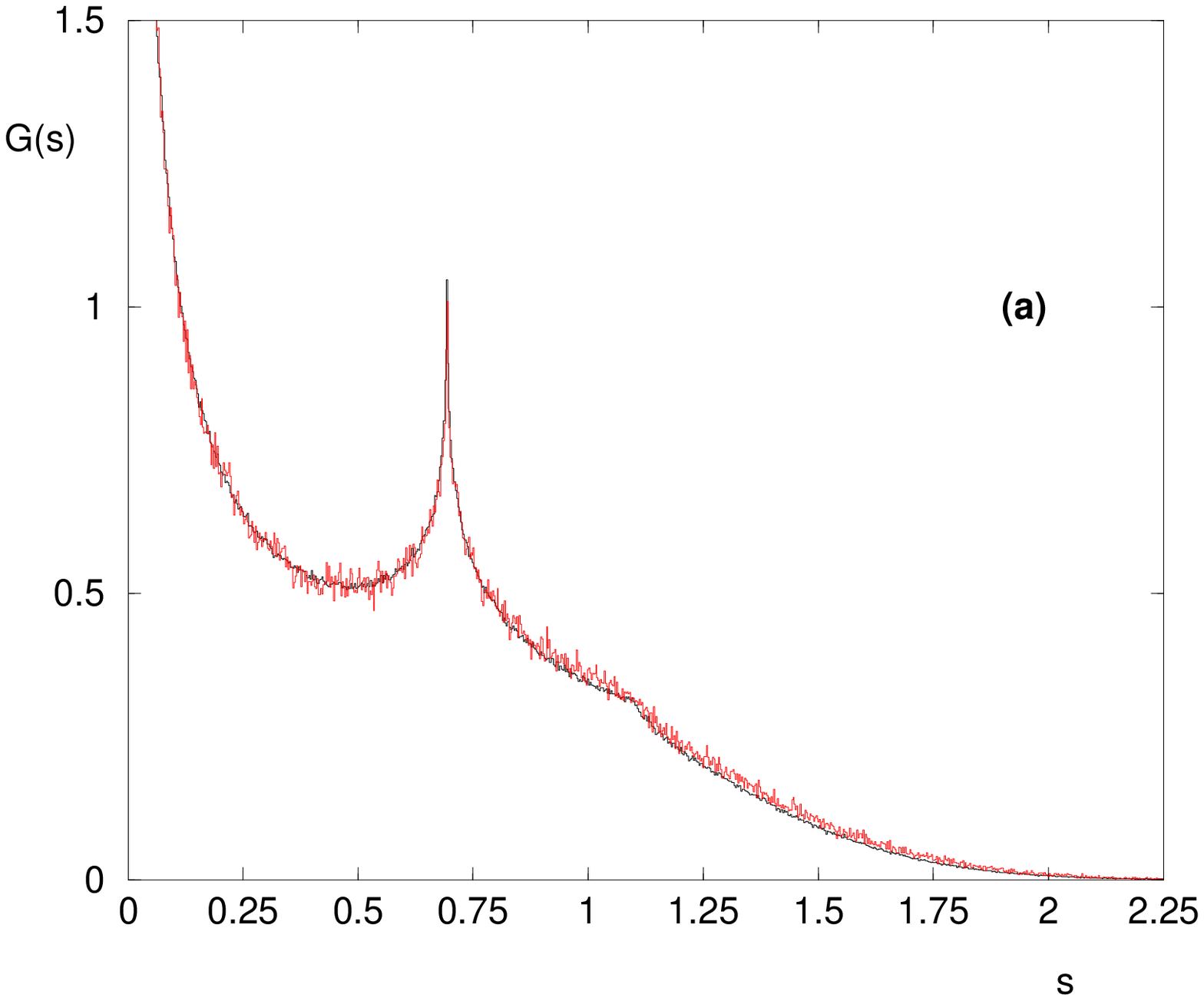}
\includegraphics[height=6cm]{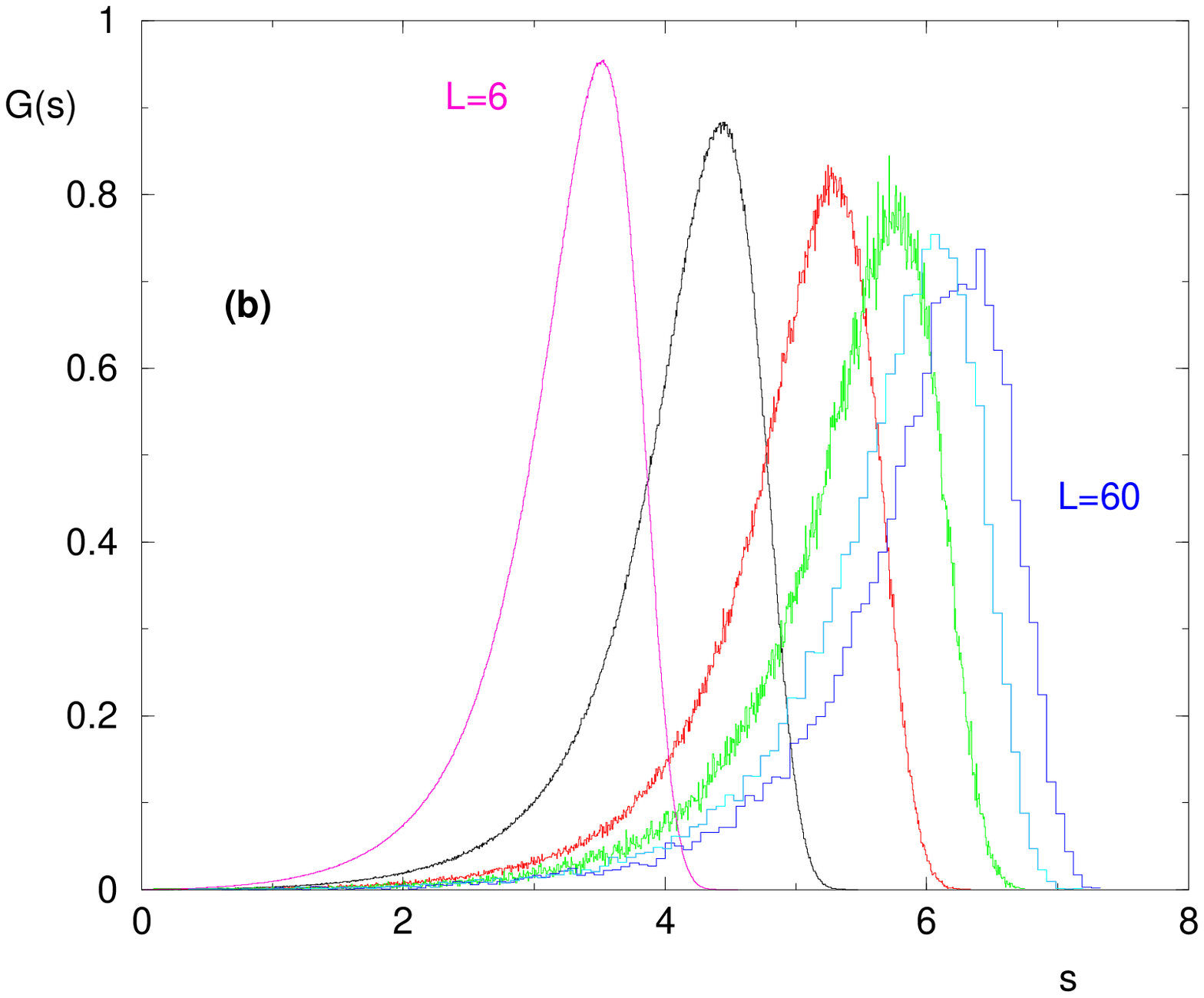}
\caption{(Color online) $d=3$
Probability distribution $G_L(s)$ of the last monomer entropy
$s_L= - \sum_{ \vec r } w_L(\vec r) \ln  w_L(\vec r)$
 (a) at  $T=0.1 <T_{gap}$  for $L=12,24$ : the Derrida-Flyvbjerg 
singularities at $s=0$, $s=\ln 2$ and $s=\ln 3$ are clearly visible.
(b) at  $T=0.6 > T_{gap}$ for $L=6,12,24,36,48,60$ : the histogram
does not reach $s=0$.   }
\label{fig3ds}
\end{figure}

The Derrida-Flyvbjerg singularities found for $T<T_{gap}$
at $w=1/n$ for the weights
translate into singularities 
in the histograms of the last-monomer entropy (Eq. \ref{entropy})
at $s=0,\ln 2, \ln 3 ...$.
In particular, the main singularity of $P_1(w)$ for $w \to 1$
 (Eq. \ref{p1singw1}) yields a corresponding singularity
at $s \to 0$
\begin{equation}
G(s) \oppropto_{s \to 0} s^{\mu(T)-1}
\label{gsing1}
\end{equation}
for $0 < T < T_{gap}$, whereas a gap $s_{min}$ appears for $T>T_{gap}$. 
This is shown on Fig. \ref{fig1ds} for $d=1$
and on Fig.  \ref{fig3ds}  for $d=3$.

\newpage

\subsection{Probability distribution $\Pi(Y_2)$
 of the parameter $Y_2$}

\begin{figure}[htbp]
\includegraphics[height=6cm]{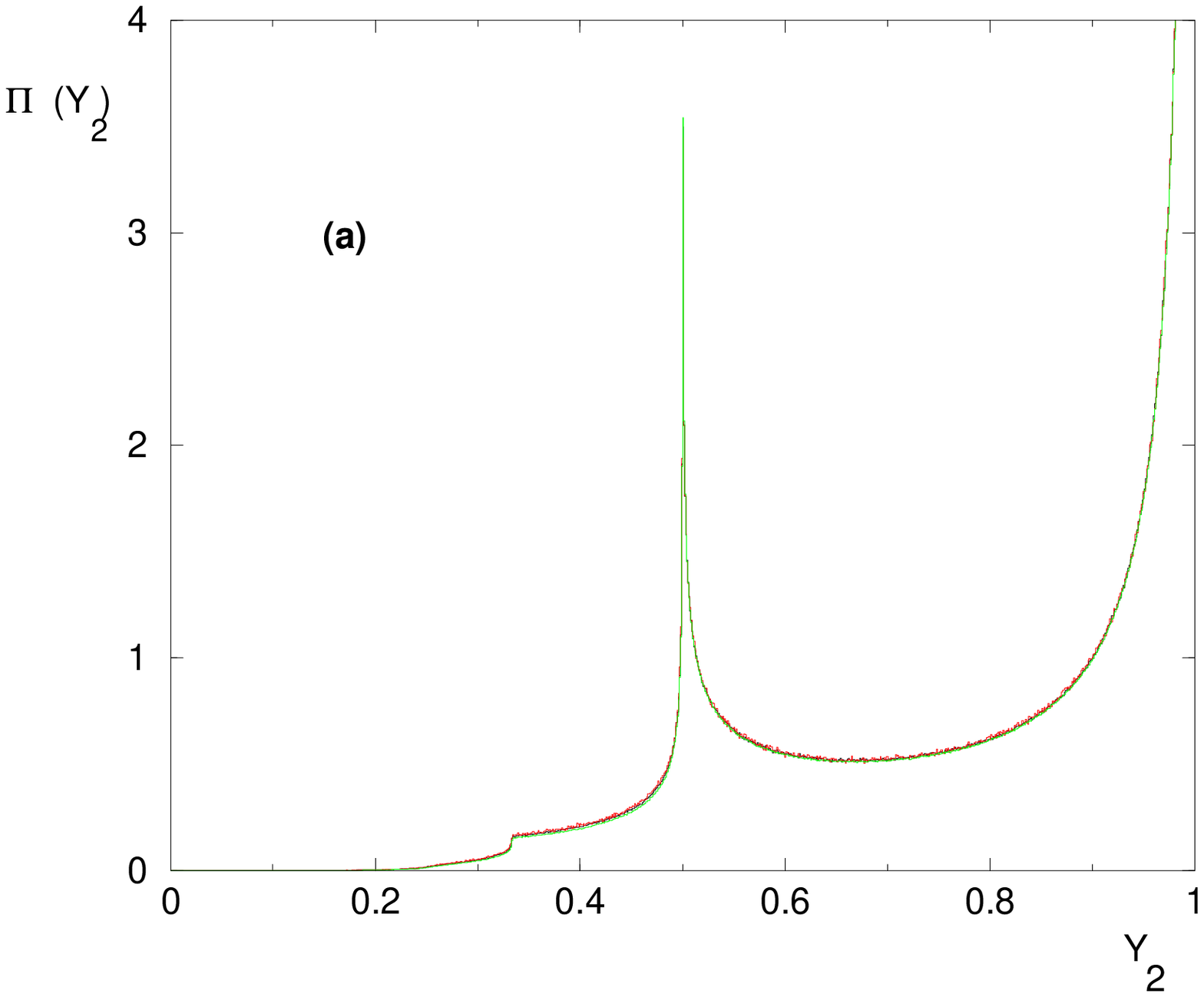}
\hspace{1cm}
\includegraphics[height=6cm]{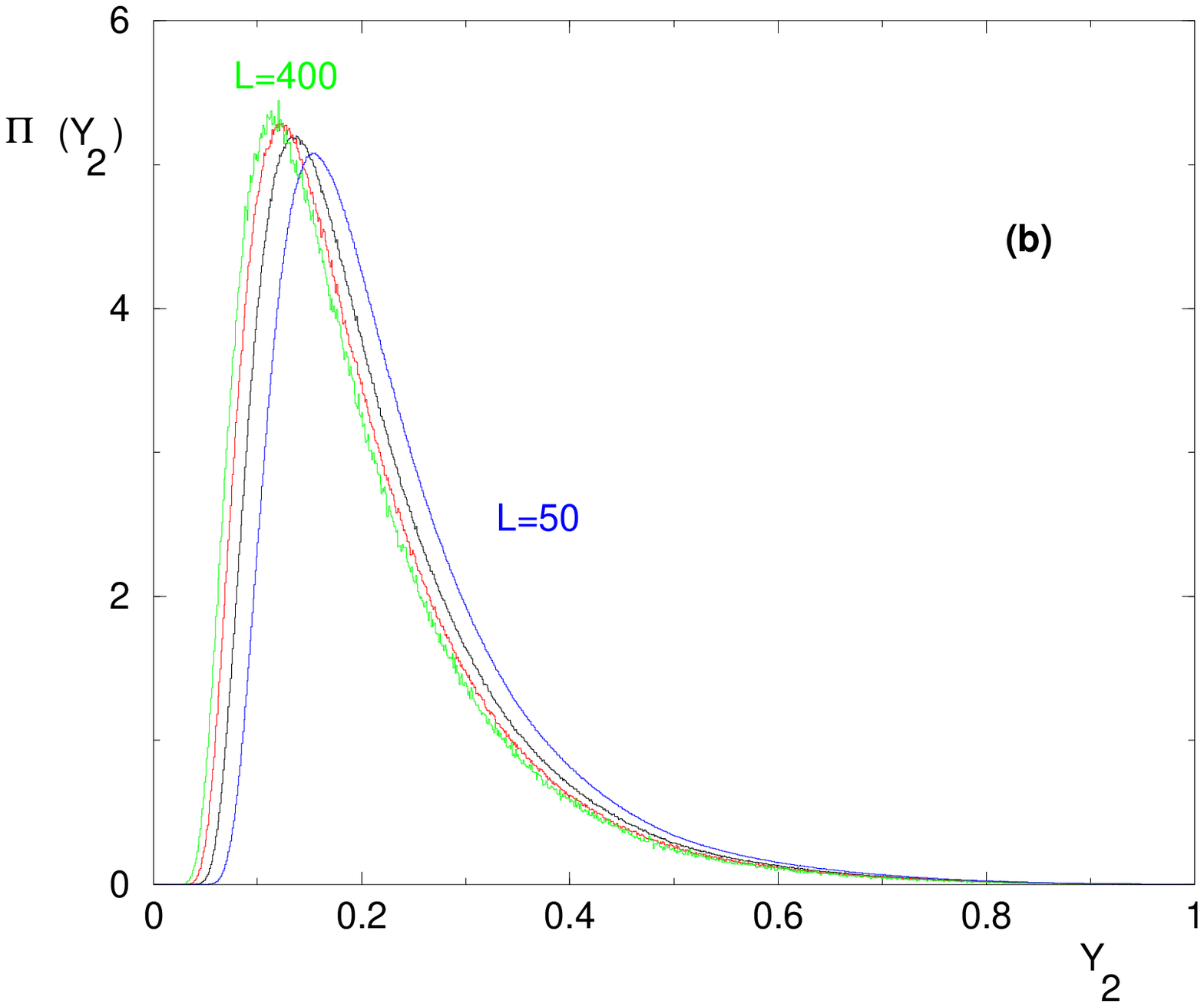}
\caption{(Color online) $d=1$ : 
Probability distribution $\Pi(Y_2)$ of the parameter
$Y_2= \sum_{ \vec r} w_L^2(\vec r)$
(a) at $T=0.1<T_{gap}$ for $L=50,100,200$ :
the Derrida-Flyvbjerg 
singularities at $Y_2=1$, $Y_2=1/2$ and $Y_2=1/3$ are clearly visible.
 (b) at $T=1.>T_{gap}$ for $L=50,100,200,400$ : the histogram
does not reach $Y_2=1$. 
  }
\label{fig1dy2}
\end{figure}

\begin{figure}[htbp]
\includegraphics[height=6cm]{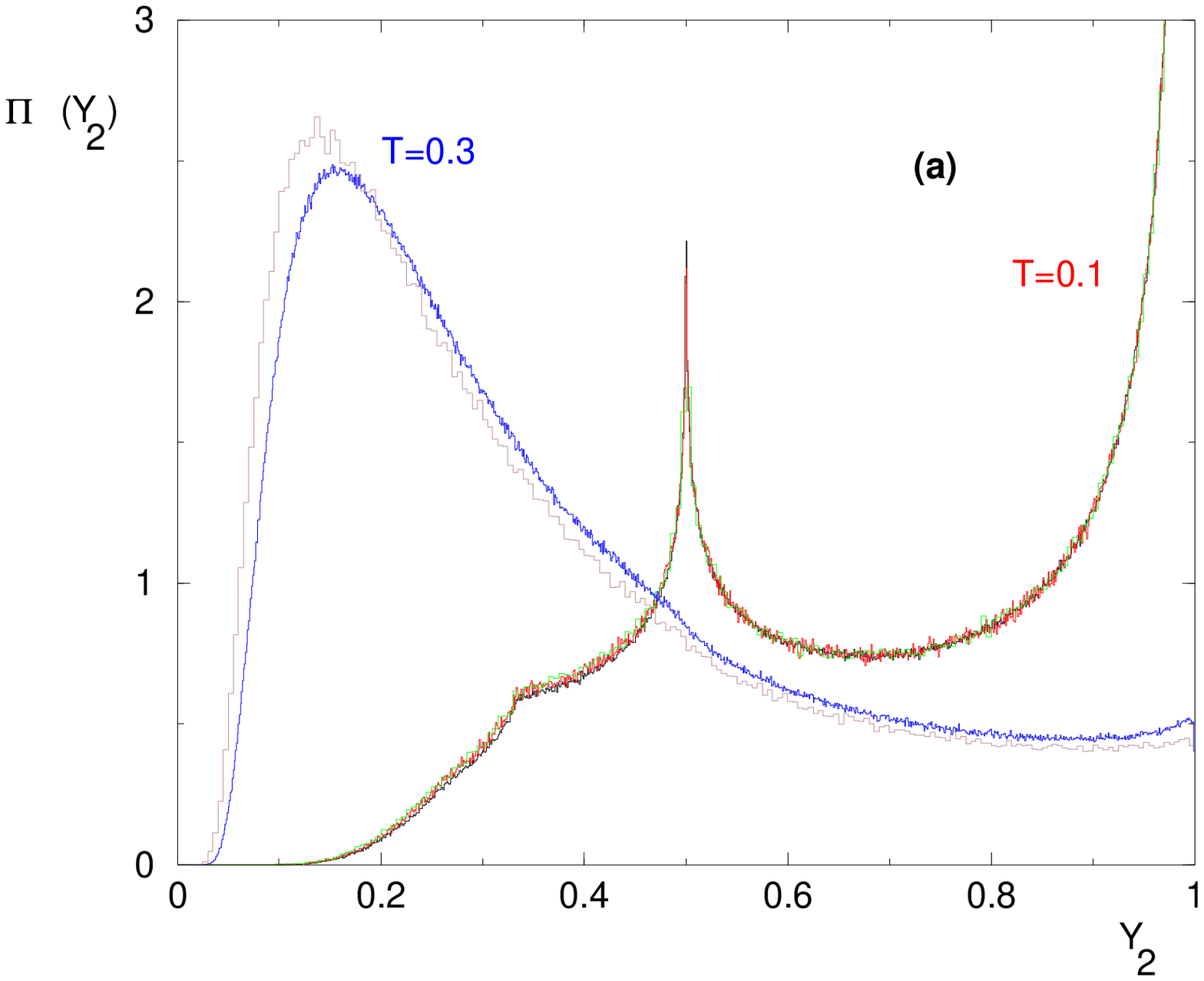}
\includegraphics[height=6cm]{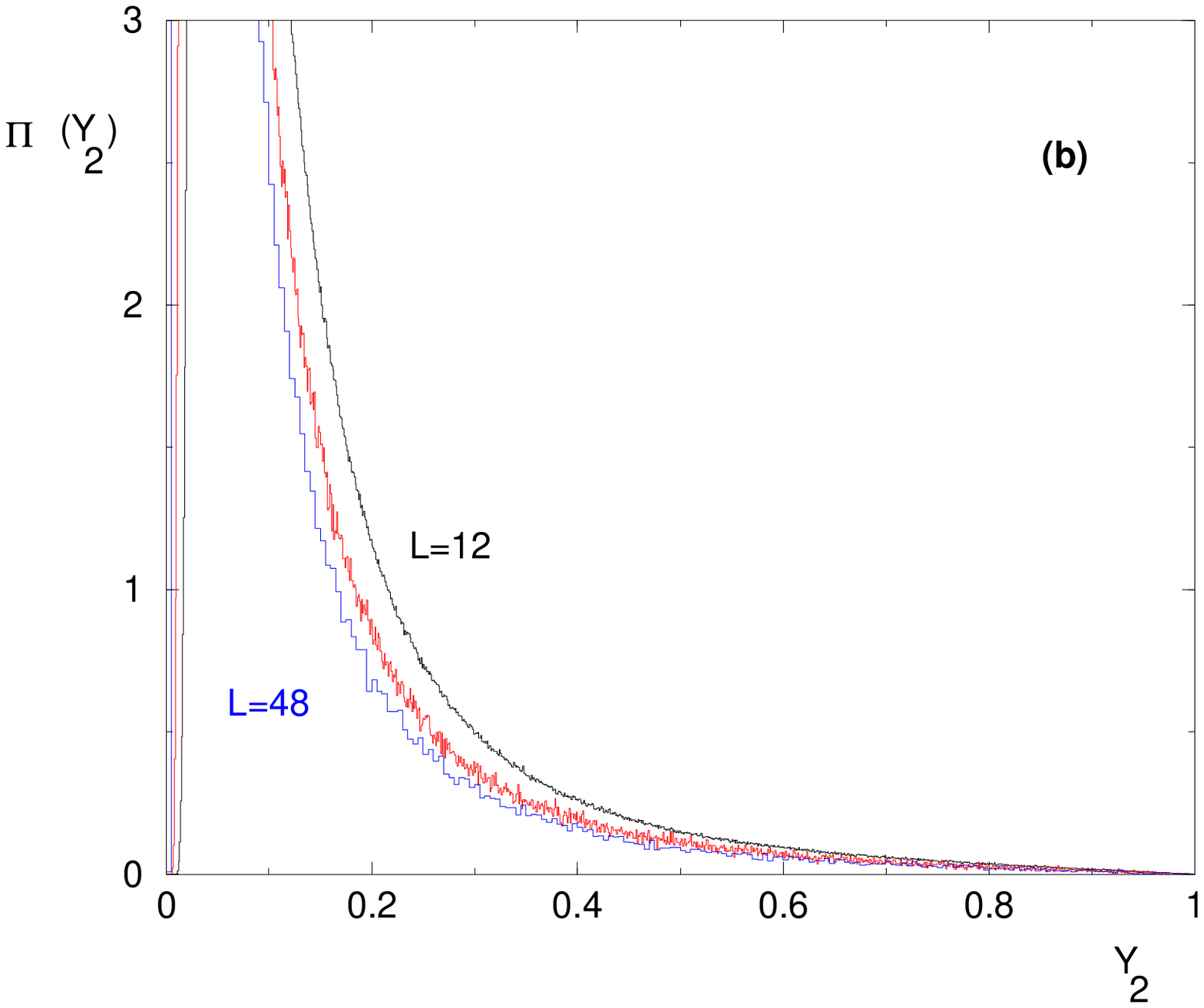}
\caption{(Color online) $d=3$  :
Probability distribution  $\Pi(Y_2)$ of the parameter
$Y_2= \sum_{ \vec r} w_L^2(\vec r)$
(a)  for $T=0.1$ ($L=12,18,24$) and $T=0.3$ ($L=12,24$) :
 the Derrida-Flyvbjerg 
singularities at $Y_2=1$, $Y_2=1/2$ and $Y_2=1/3$ are clearly visible.
(b) for $T=0.5 \sim T_{gap}$ with  $L=12,24,48$    }
\label{fig3dy2}
\end{figure}

The parameter $Y_2$ defined in Eq. \ref{y2def} can reach the value $Y_2 \to 1$
only if the maximal weight $w^{max}$ also reaches $w^{max} \to 1$.
As a consequence, the probability distribution $\Pi(Y_2)$
has the same singularity near $Y_2\to 1$ as in Eq. (\ref{p1singw1})
\begin{equation}
\Pi(Y_2) \oppropto_{Y_2 \to 1} (1-Y_2)^{\mu(T)-1}
\label{pising1}
\end{equation}
for $0 < T < T_{gap}$. 
Beyond this main singularity, 
the distribution
$\Pi(Y_2)$ presents the characteristic Derrida-Flyvbjerg 
singularities at $Y_2=1/n$ as shown on Fig. \ref{fig1dy2} a for $d=1$
and on Fig. \ref{fig3dy2} a for $d=3$.
Again for $T>T_{gap}$, a gap appears as 
 shown on Fig. \ref{fig1dy2} b for $d=1$
and on Fig. \ref{fig3dy2} b for $d=3$.

\subsection{Density $f(w)$ }

\begin{figure}[htbp]
\includegraphics[height=6cm]{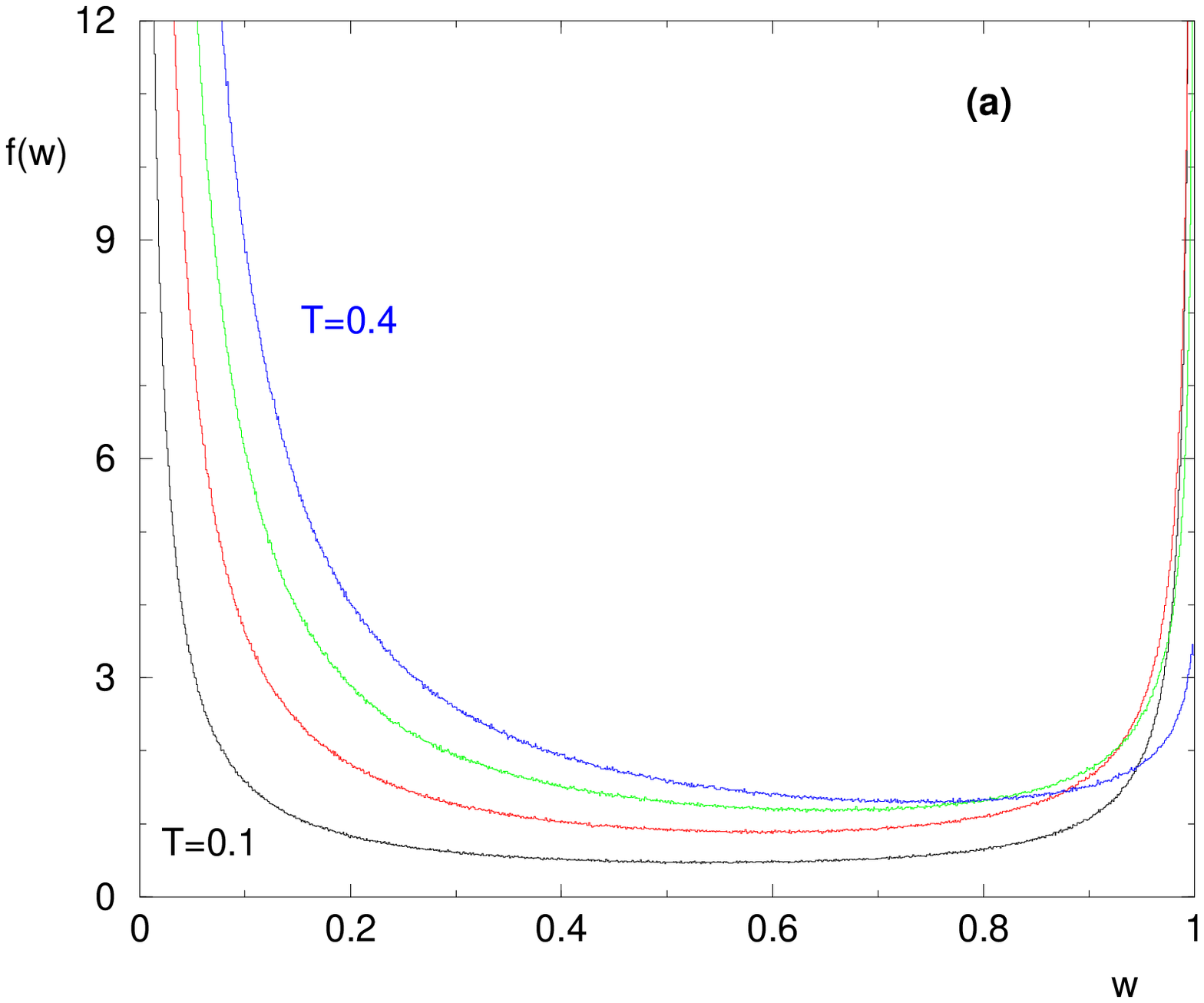}
\hspace{1cm}
\includegraphics[height=6cm]{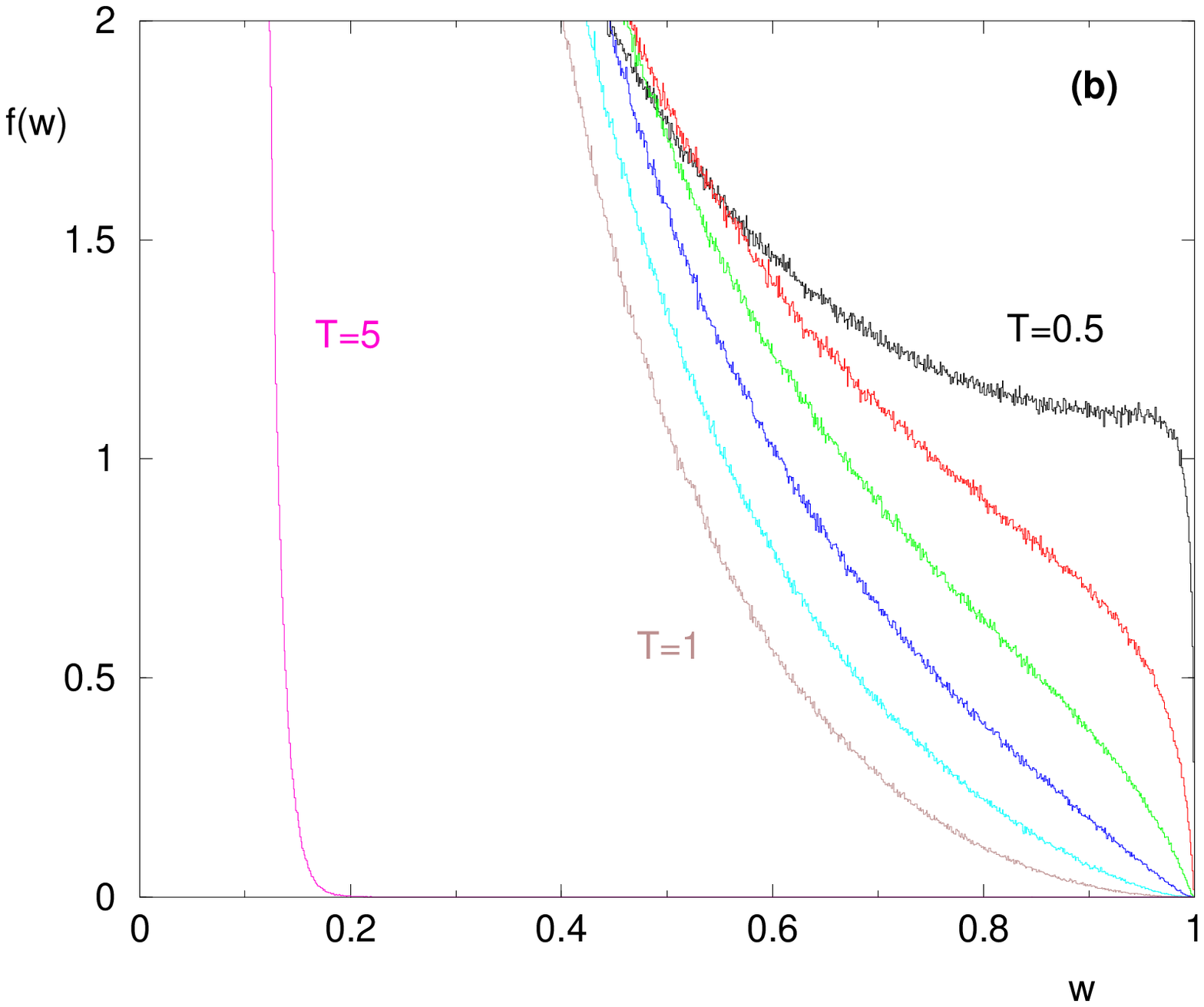}
\caption{(Color online) $d=1$ : 
Weight density  $f(w)$
(a) at $T=0.1,0.2,0.3,0.4$ where $\mu(T)<1$ for $L=200$ :
 the weight density $f(w)$ diverges at $w \to 0$ and $w \to 1$.
 (b) at $T=0.5,0.6,0.7,0.8,0.9,1.,5.$  for $L=200$ :
these curves show the temperature dependent singularity at $w=1$.  }
\label{fig1df}
\end{figure}

\begin{figure}[htbp]
\includegraphics[height=6cm]{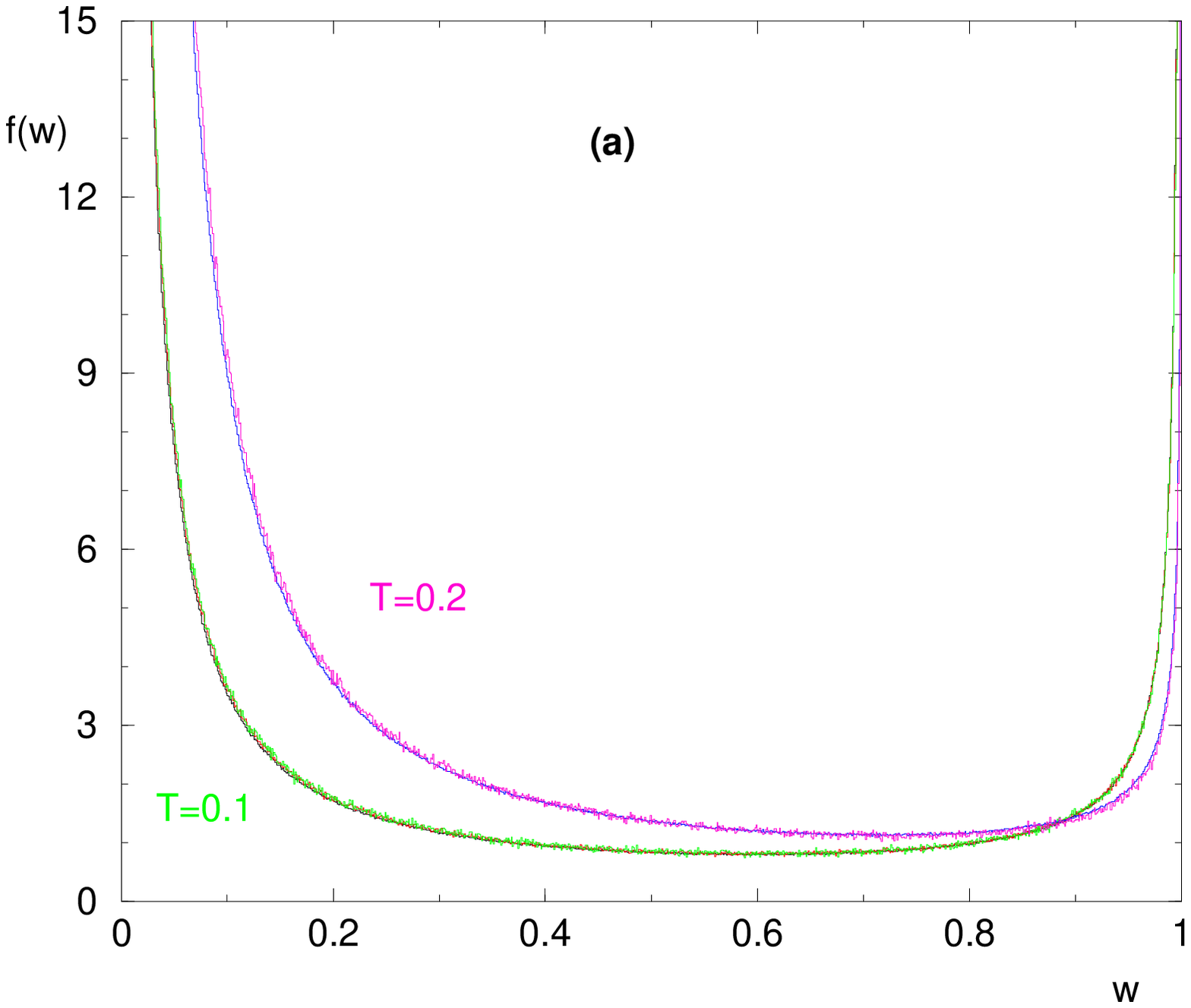}
\hspace{1cm}
\includegraphics[height=6cm]{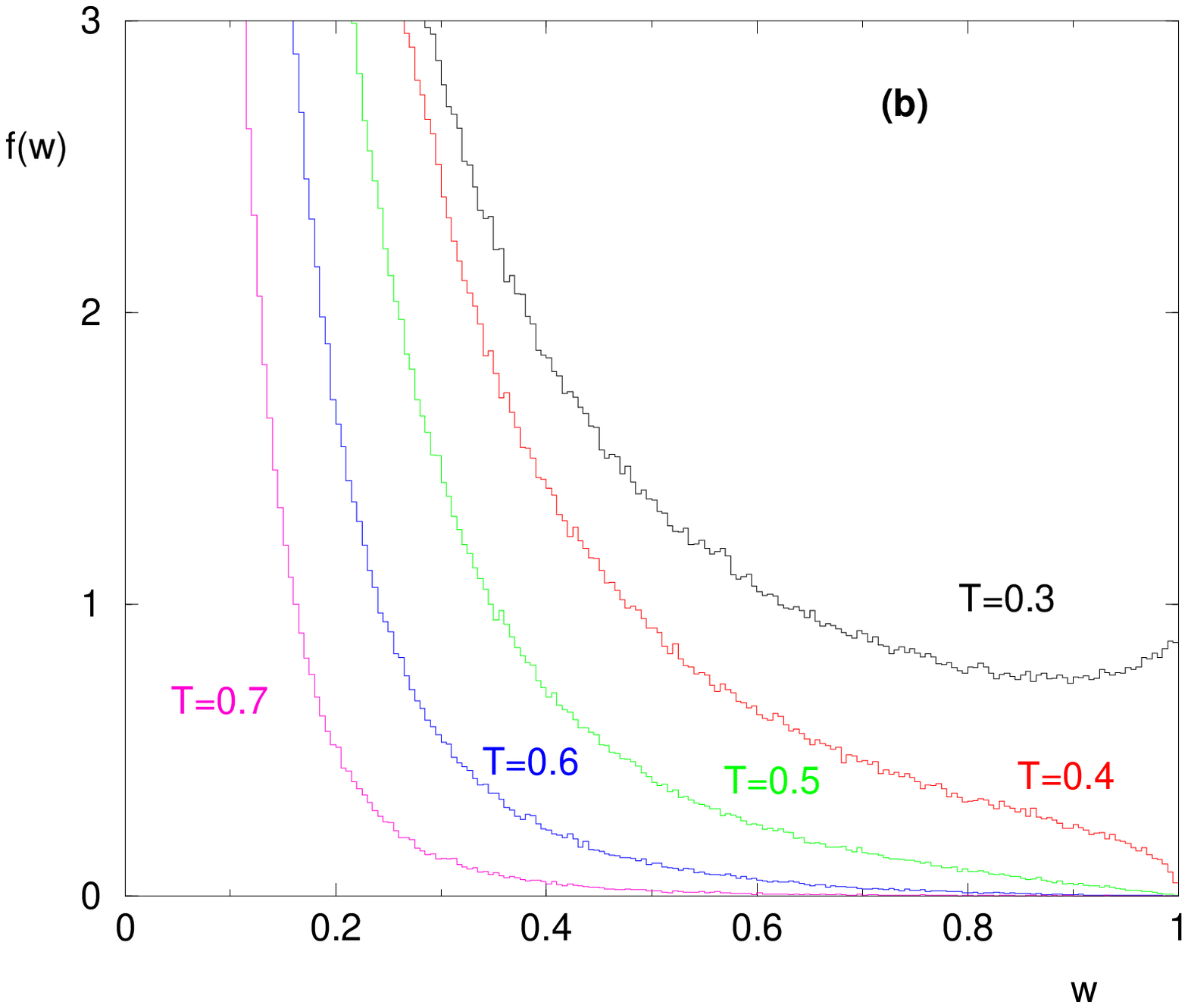}
\caption{(Color online) $d=3$  :
Weight density  $f(w)$
(a) 
 for $T=0.1$ ($L=12,18,24$) and $T=0.2$ ($L=12,24$) where $\mu(T)<1$ :
  the weight density $f(w)$ diverges at $w \to 0$ and $w \to 1$.
(b)  for $T=0.3,0.4,0.5,0.6,0.7$ ($L=24$) :
these curves show that the gap appears around $T_{gap}(d=3) \sim 0.5$   }
\label{fig3df}
\end{figure}

The density $f(w)$ introduced in Eq. (\ref{densityfdef}) is shown on
Fig. \ref{fig1df} and Fig. \ref{fig3df}
for $d=1$ and $d=3$ respectively.

By construction, this density coincides with the maximal weight
distribution $P_1(w)$ for $w>1/2$,
with the sum ($P_1(w)+P_2(w)$) of the two largest weight distributions
 for $1/3<w<1/2$, and so on \cite{Der_Fly}.
As a consequence, $f(w)$ has the same singularity near 
$w \to 1$ as $P_1(w)$ (Eq. \ref{p1singw1} ),
and the same gap (Eq. \ref{p1gapw0} ) as long as $w_0(T)>1/2$.
The only other singularity is near $w \to 0$ where $f(w)$ diverges 
in a non-integrable manner, because in the $L \to \infty$, there is an
infinite number of vanishing weights (only the product $(w f(w))$ has
to be integrable at $w=0$ as a consequence of the normalization
condition of Eq. \ref {normay1}).

\subsection{ Moments  $\overline{Y_k} $ }

\begin{figure}[htbp]
\includegraphics[height=6cm]{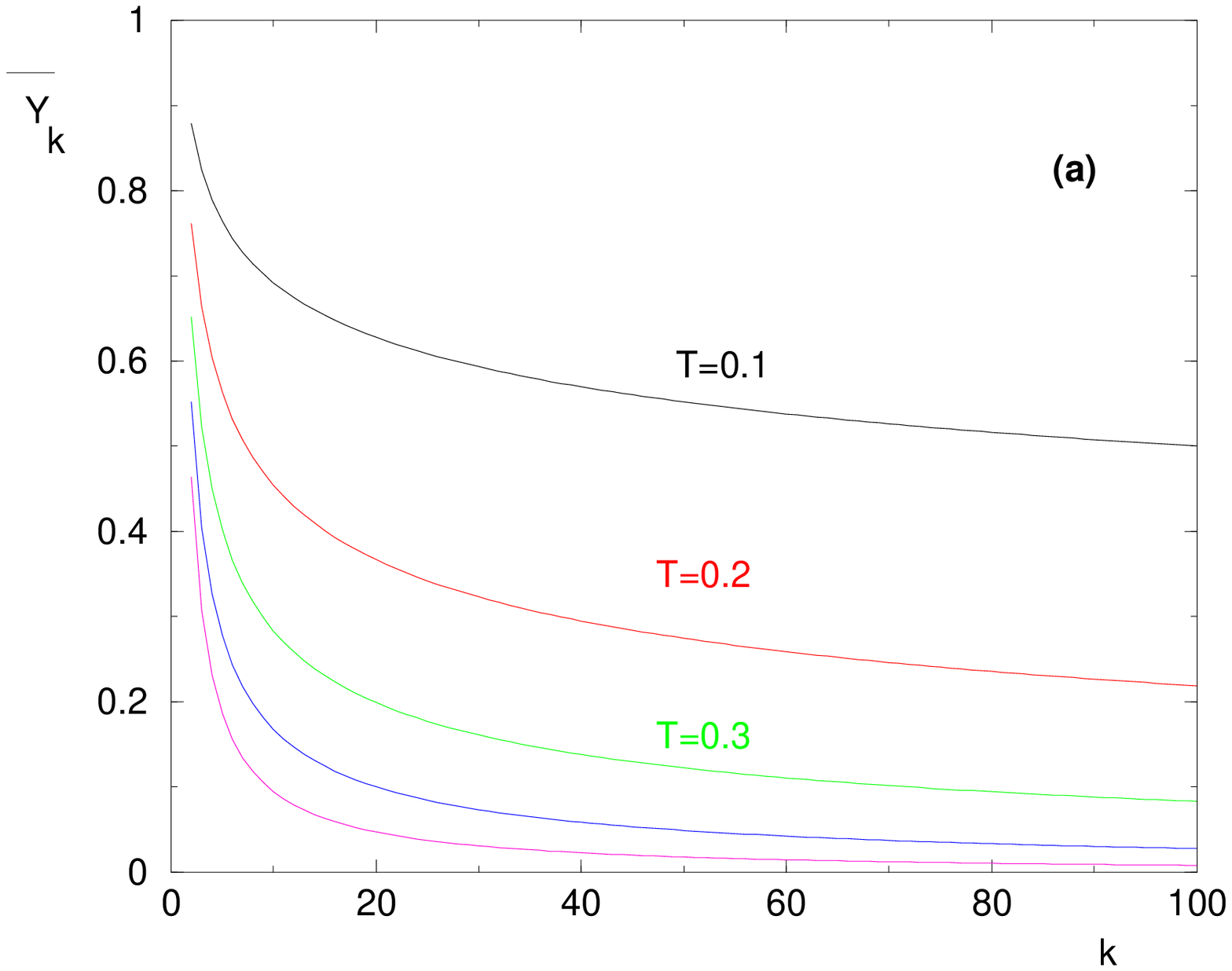}
\hspace{1cm}
\includegraphics[height=6cm]{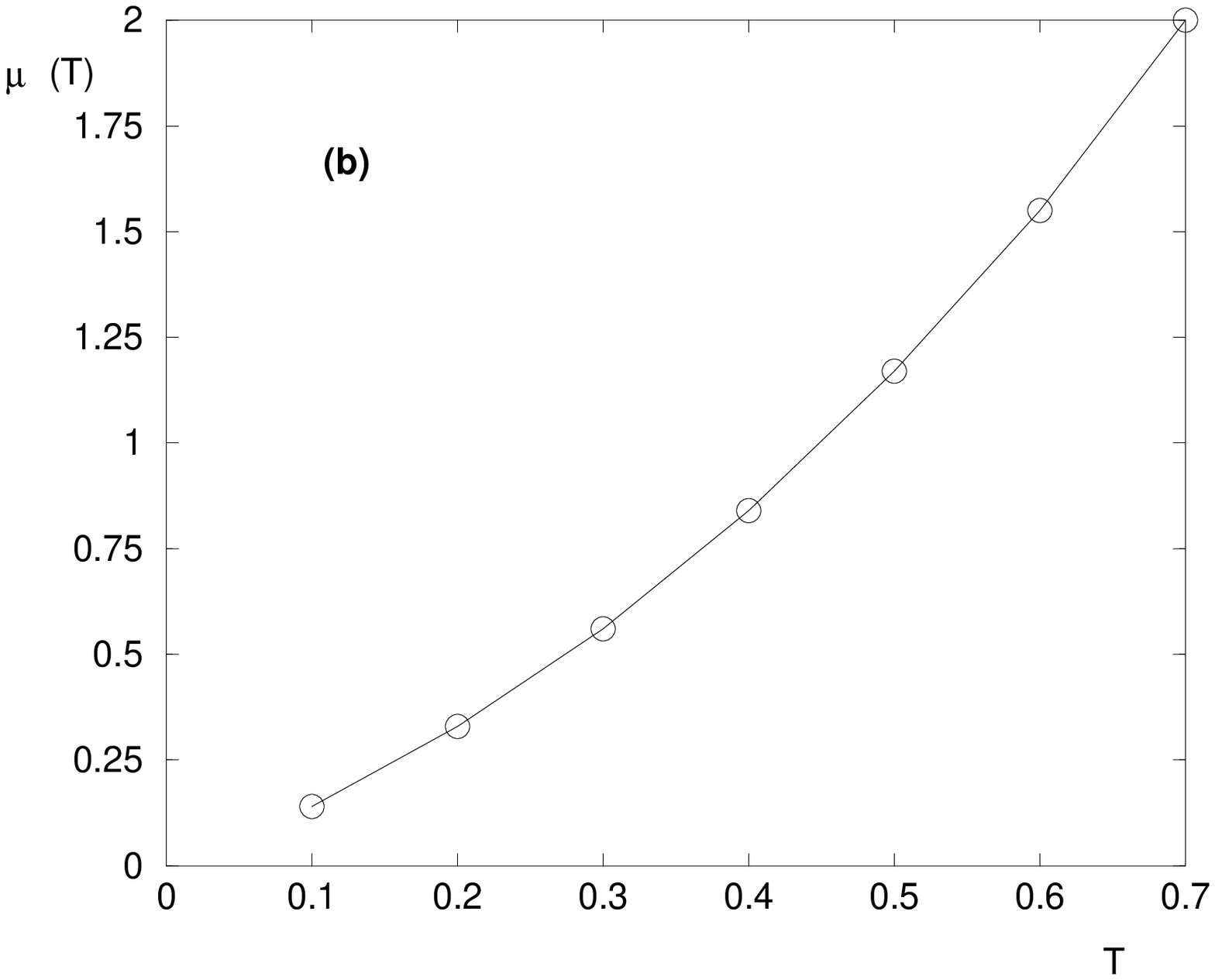}
\caption{(Color online) $d=1$ : 
(a) Decay of the moments $\overline{Y_k} $ of Eq. \ref{ykdecay}
as a function of $k \leq 100$ for $L=800$ and
 $T=0.1,0.2,0.3,0.4,0.5$
(b)
Exponent $\mu(T)$ as measured from the slope of the log-log decay
in the asymptotic region.}
\label{fig1dmu}
\end{figure}

\begin{figure}[htbp]
\includegraphics[height=6cm]{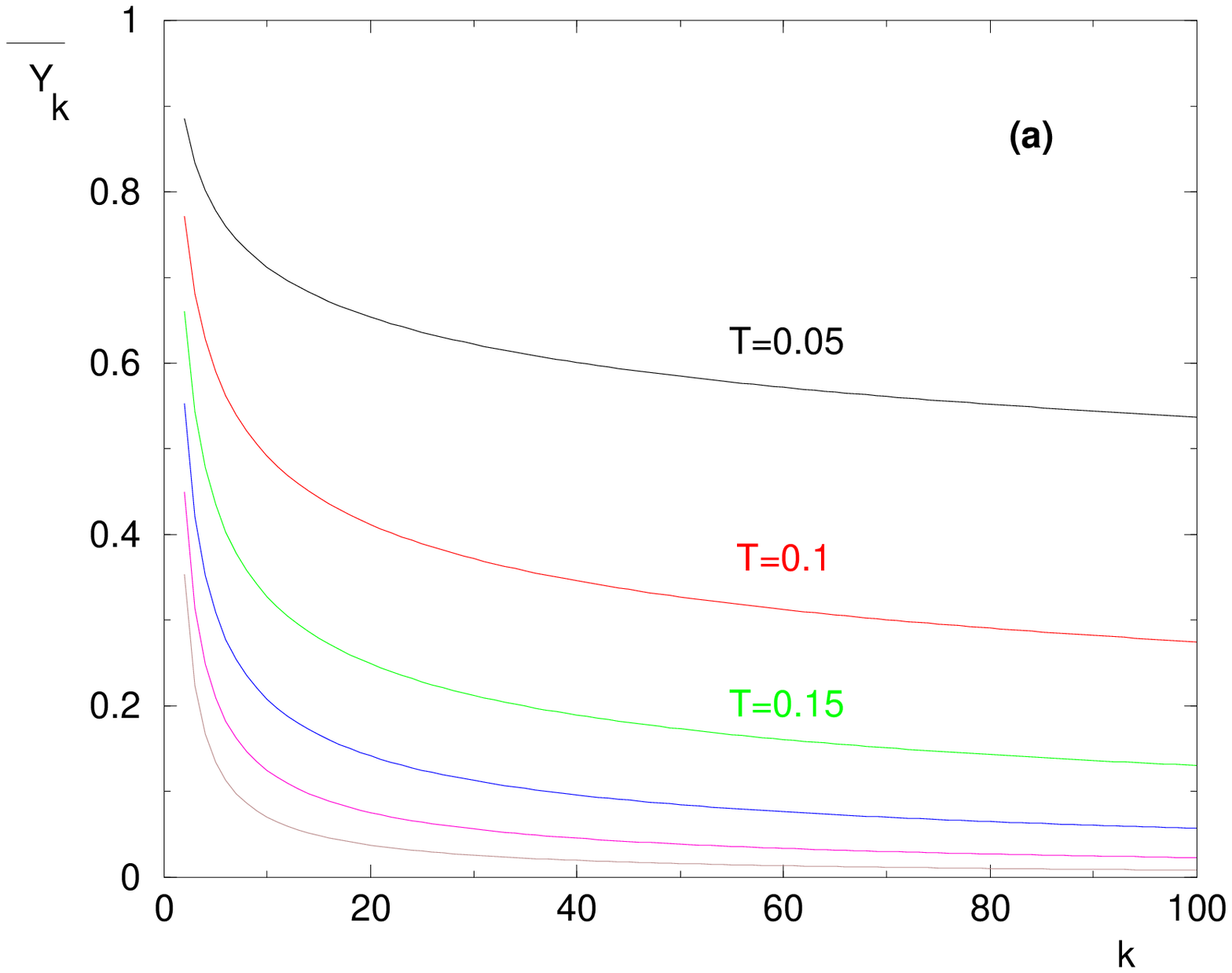}
\hspace{1cm}
\includegraphics[height=6cm]{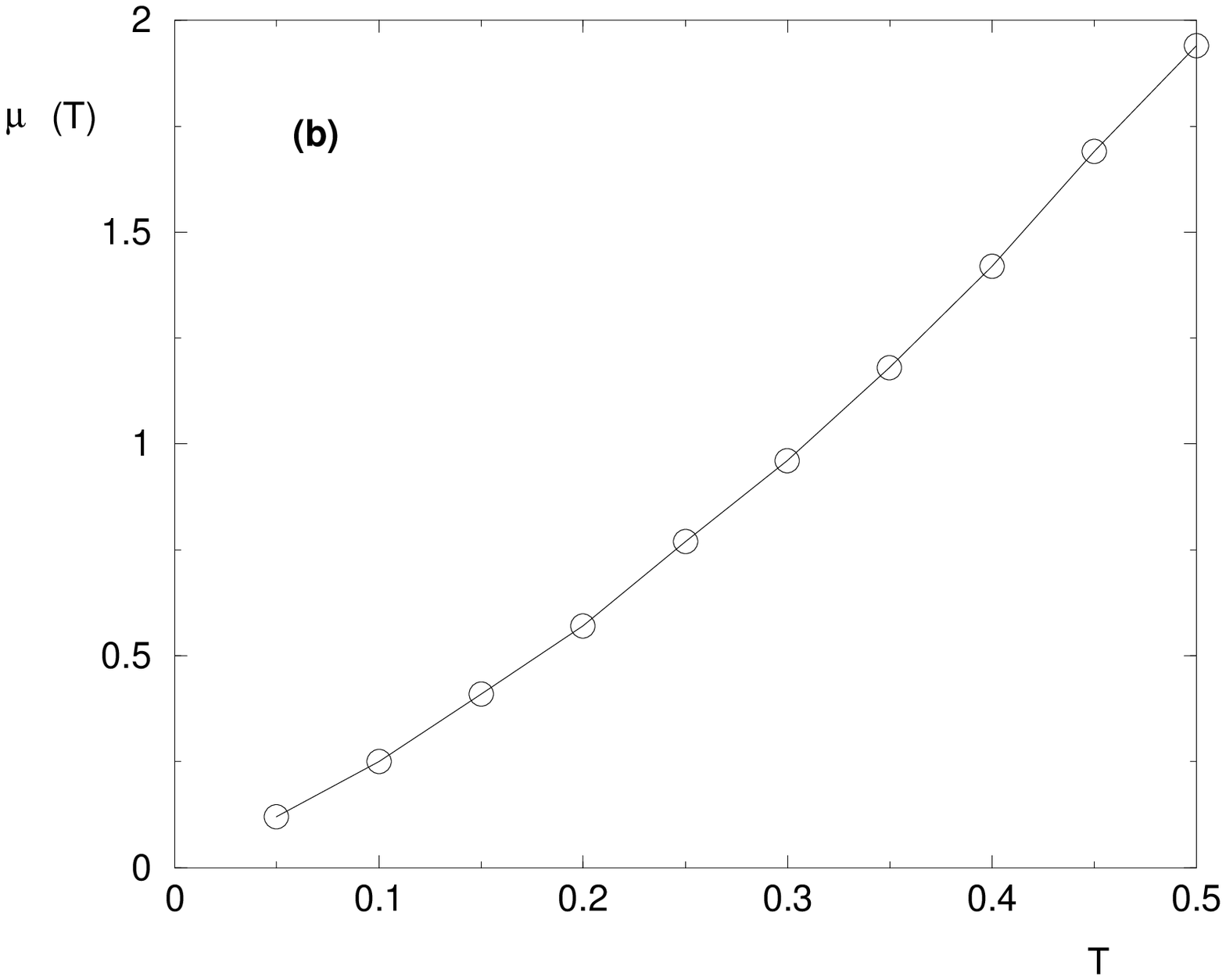}
\caption{(Color online) $d=3$  :
(a) Decay of the moments $\overline{Y_k} $ of Eq. \ref{ykdecay}
as a function of $k \leq 100$ for $L=48$ and
 $T=0.05,0.1,0.15,0.2,0.25,0.3$
(b)
Exponent $\mu(T)$ as measured from the slope of the log-log decay
in the asymptotic region.    }
\label{fig3dmu}
\end{figure}

For $0<T\leq T_{gap}$, where $P_1(w)$ and $f(w)$ behaves near
 $w \to 1$ as in Eq. \ref {p1singw1},
the decay in $k$ of the averaged moments $\overline{Y_k(i)}$ (Eq. \ref{ykdef})
 follow a power-law of exponent $\mu(T)$
\begin{equation}
\overline{Y_k(i)} \oppropto_{k \to \infty}  \frac{1}{k^{\mu(T)}} 
 \  \  {\rm for } \ \ T \leq T_{gap}
\label{ykdecay}
\end{equation}
The behavior of $\overline{Y_k}$ for $k \leq 100$ are shown on
Fig. \ref{fig1dmu} a for $d=1$ and Fig. \ref{fig3dmu} a for $d=3$.
The corresponding exponent $\mu(T)$ are shown on
Fig. \ref{fig1dmu} b for $d=1$
and Fig. \ref{fig3dmu} b for $d=3$.

For $T>T_{gap}$ where there exists a gap $w_0(T)$ for $P_1(w)$, 
the behavior of Eq. \ref{p1gapw0}
also applies to $f(w)$ as long as $w_0(T)>1/2$ 
(since $f(w)=P_1(w)$ for $w>1/2$ as mentioned above)
and thus the decay is then exponential
\begin{equation}
\overline{Y_k(i)} \oppropto_{k \to \infty}  \frac{
(w_0(T))^k}{k^{1+\sigma}} 
 \  \  {\rm for } \ \ T > T_{gap}
\label{ykgap}
\end{equation}

\section{  Study of spatial properties }

\label{spatial} 

In the previous Section, we have studied in details the statistics of the weights
independently of their spatial organization.
In this section, we study the statistics
of the transverse spatial correlation 
\begin{equation}
C( r ) = < w(\vec r_{pref}) w(\vec r_{pref} + \vec r ) >
\label{corre}
\end{equation}
centered on the preferred
position
$\vec r_{pref}$ of maximal weight (Eq. \ref{wmax}).
We first recall the predictions of the 
droplet scaling analysis \cite{Fis_Hus,Hwa_Fis}
that will be useful to analyse our numerical results.

\subsection{ Reminder on the droplet scaling analysis}

The droplet theory for directed polymers \cite{Fis_Hus,Hwa_Fis},
is very similar to the droplet theory of spin glasses \cite{Fis_Hus_SG}.
It is a scaling theory that can be summarized as follows.

\subsubsection{Statistics of low energy excitations above the
ground state }

At very low temperature $ T \to 0$, all observables are governed by
the statistics of low energy excitations above the ground state.
An excitation of large length $l$ costs a random energy
\begin{eqnarray}
 \Delta E(l) \sim l^{\theta} u
\label{ground}
\end{eqnarray}
where $u$ is a positive random variable distributed with some law $Q_0 (u)$
having  some finite density at the origin  $Q_0 (u=0) >0$.
The exponent $\theta$ is the exponent governing 
the fluctuation of the energy of the ground state
is exactly known in one-dimension
$\theta(d=1)=1/3$ \cite{Hus_Hen_Fis,Kar,Joh,Pra_Spo}
and for the mean-field version on the Cayley tree
 $\theta(d=\infty)=0$ \cite{Der_Spo}.
In finite dimensions $d=2,3,4,5,...$, the exponent $\theta(d)$ has
been numerically measured, and we only quote here the results of the
most precise study we are aware of \cite{Mar_etal} for dimensions
$d=2,3$ : $\theta(d=2)=0.244$ and $\theta(d=3)  = 0.186$.

From (\ref{ground}), the probability distribution of 
large excitations $ l \gg 1$ reads within the droplet theory
\begin{eqnarray}
dl \rho (E=0,l)  \sim \frac{ dl }{l} e^{- \beta \Delta E(l)} 
\sim  \frac{ dl }{l} e^{- \beta l^{\theta} u  }
\label{rhodroplet}
\end{eqnarray}
where the factor $dl/l$ comes from the notion of independent excitations
\cite{Fis_Hus_SG}. In particular, its average over the disorder
follows the power-law
\begin{eqnarray}
dl \overline{ \rho (E=0,l) }  
\sim \int_0^{+\infty} du Q_0(u)  \frac{ dl }{l} e^{- \beta l^{\theta} u  }
= T Q(0) \frac{ dl }{l^{1+\theta}}
\label{rhoavt0}
\end{eqnarray}
This prediction describes very well the numerical data in 
the regime $1 \ll l \ll L$ in dimensions $d=1,2,3$ \cite{DPexcita}.

Since correlation functions at large distance are directly
related to the probability of large excitations,
we already see that the low temperature phase
is very non-trivial from the point of view of correlations
lengths : the typical exponential decay (\ref{rhodroplet}) indicates
a finite typical correlation length $\xi_{typ}(T)$,
whereas the averaged power-law behavior (\ref{rhoavt0}) means
that the averaged correlation length $\xi_{av}(T)$
is actually infinite in the whole low temperature phase
\begin{eqnarray}
\xi_{av}(0<T \leq T_c) =\infty
\end{eqnarray}
 Note that within the droplet theory of 
spin-glasses \cite{Fis_Hus_SG}, the correlation length $\xi_{av}(T)$
is also infinite in the whole low temperature phase
for the same reasons.

\subsubsection{Low temperature phase governed by a zero-temperature
fixed point}

According to the droplet scaling theory \cite{Fis_Hus,Hwa_Fis}
 the whole low temperature phase $0<T<T_c$
is governed by a zero-temperature fixed point. 
However, many subtleties arise because the temperature
is actually `dangerously irrelevant'. 
The main conclusions of the droplet analysis \cite{Fis_Hus,Hwa_Fis}
can be summarized as follows.
The scaling (\ref{ground}) governs the free energy cost
of an excitation of length $l$, provided one introduces
a longitudinal correlation length $\xi_{//}(T)$ to rescale the length $l$
\begin{eqnarray}
\Delta F (l ) = \left( \frac{l}{\xi_{//}(T) } \right)^{\theta} u
\label{deltaF}
\end{eqnarray}
Here as before, $u$ denotes
 a positive random variable distributed with some law $Q (u)$
having  some finite density at the origin  $Q (u=0) >0$.
As a consequence, 
the probability of a droplet of size $l \gg 1$ follows the scaling
form
\begin{equation}
dl \rho(l) \sim \frac{dl}{l} 
e^{- u \left( \frac{l}{\xi_{//}(T)} \right)^{\theta} }
\label{droplet}
\end{equation}
In particular, the typical behavior follows an exponential decay with exponent $\theta$
\begin{equation}
\overline{ \ln \rho(l) } \sim - \left( \frac{l}{\xi_{//}(T)} \right)^{\theta}
\int_0^{+\infty} du \  u Q(u)
\label{rhotyp}
\end{equation}
whereas the average over the disorder
follows the power-law
\begin{eqnarray}
dl \overline{ \rho (l) }  
\sim  Q(0) \frac{ dl }{l} \left( \frac{\xi_{//}(T)}{l} \right)^{\theta}
\label{rhoav}
\end{eqnarray}
This average is governed by the rare events having $u \sim 0$.

A droplet of longitudinal size $l$ corresponds 
a transverse distance 
$ r \sim l^{\zeta}$, where
\begin{equation}
\zeta= \frac{1+\theta}{2}
\label{zeta}
\end{equation}
is the roughness exponent of the low temperature phase \cite{Hal_Zha}.
Via the change of variable $ r \sim l^{\zeta}$, the droplet
distribution of Eq. \ref{droplet} translates into the following
scaling form for the correlation at large distance $r$ 
\cite{Fis_Hus,Hwa_Fis}
\begin{equation}
dr r^{d-1} C( r ) = dl \rho(l)
 = \frac{dr}{r} e^{- u 
\left( \frac{r}{\xi_{\perp}(T)} \right)^{\frac{\theta}{\zeta}} }
\label{change}
\end{equation}
where the transverse correlation length 
reads $\xi_{\perp}(T) \sim [\xi_{//}(T)]^{\zeta}$, i.e. finally
\begin{equation}
 C( r ) = 
  \frac{1}{r^d} e^{- u 
\left( \frac{r}{\xi_{\perp}(T)} \right)^{\frac{\theta}{\zeta}} }
\label{correlationLinfty}
\end{equation}
As a consequence, the typical behavior 
follows an exponential decay with exponent $\theta/\zeta$
\begin{equation}
\overline{ \ln C( r ) } \sim - 
\left( \frac{r}{\xi_{\perp}(T)} \right)^{\frac{\theta}{\zeta}} -d \ln r
\label{ctypLinfty}
\end{equation}
whereas the average over the disorder is governed by the rare events
and follows the power-law
\begin{equation}
 \overline{ C( r )} \sim \frac{1}{r^d} 
\left( \frac{\xi_{\perp}(T)}{r} \right)^{\frac{\theta}{\zeta}}
\label{cavLinfty}
\end{equation}

We now describe our numerical data and compare with
these predictions.

\subsection{ Disorder averaged correlation $\overline{ C_L( r )}$  }

\begin{figure}[htbp]
\includegraphics[height=6cm]{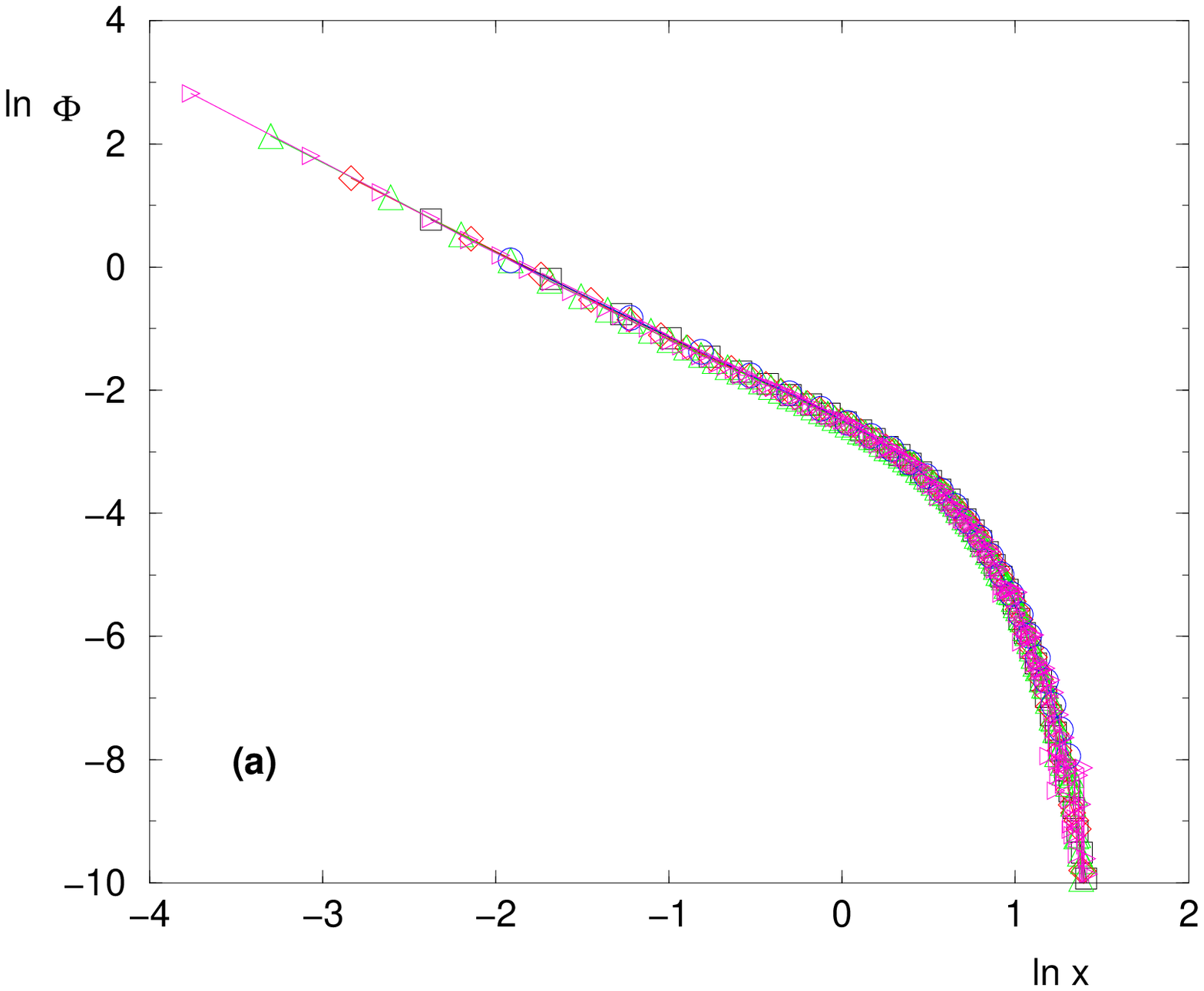}
\hspace{1cm}
\includegraphics[height=6cm]{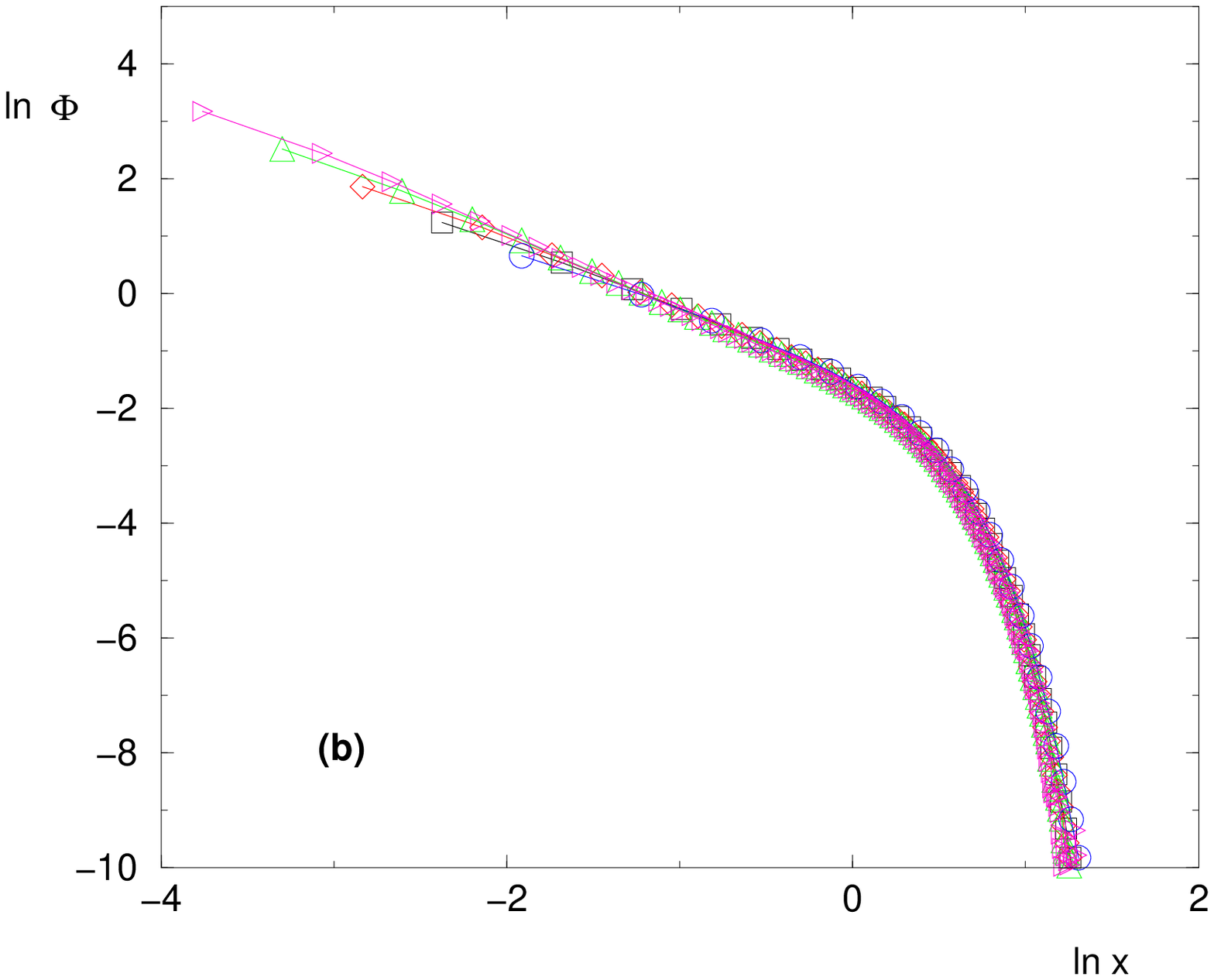}
\caption{(Color online) 
Disorder averaged correlation $\overline{ C_L( r )}$ in $d=1$ :
finite-size scaling of (Eq. \ref{cavfss}) : 
 $\ln \Phi = \ln(L \overline{ C_L( r )})$ as a  function of $\ln x= \ln(r/L^{2/3})$
 for $L=50 (\bigcirc),100 (\square),200 (\Diamond),400(\triangle),800 (\rhd)$  
(a) for $T=0.2<T_{gap}$   
 (b) for $T=1. > T_{gap}$           }
\label{fig1dcav}
\end{figure}

\begin{figure}[htbp]
\includegraphics[height=6cm]{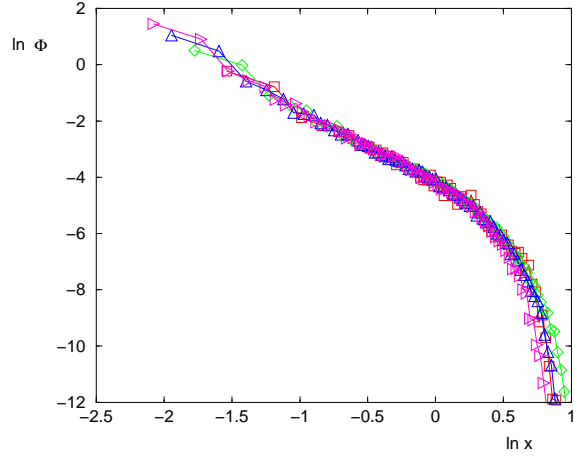}
\caption{(Color online) $d=3$ :
Disorder averaged correlation $\overline{ C_L( r )}$ in $d=3$
finite-size scaling of (Eq. \ref{cavfss}) : (Eq. \ref{cavfss}) :
 $\ln \Phi = \ln(L^{d \zeta + \theta} \overline{ C_L( r )})$
 as a  function of $\ln x = \ln (r/L^{\zeta})$
for $T=0.1$ $L=24 (\square), 36 (\Diamond),48(\triangle) ,60(\rhd) $
.}
\label{fig3dcavt0.1}
\end{figure}

The previous predictions concern the limit $L \to \infty$.
To compare with our numerical data, we now recall the
corresponding finite-size behaviors within the droplet theory
\cite{Hwa_Fis}.
The power-law behavior (Eq. \ref{cavLinfty})
for the averaged correlation translates into
the following scaling form for finite $L$
\begin{equation}
 \overline{ C_L( r )} \sim \frac{1}{L^{d \zeta+\theta} } 
\Phi \left( x=\frac{r}{L^{\zeta} } \right)
\label{cavfss}
\end{equation}
where the scaling function $\Phi(v)$ behaves as the following
power-law 
\begin{equation}
\Phi(v) \oppropto_{v \to 0} \frac{1}{v^{d +\frac{\theta}{\zeta}}}
\label{phiv}
\end{equation}
 to recover Eq. \ref{cavLinfty} as $L \to \infty$.

We show on Fig. \ref{fig1dcav} the finite-size scaling analysis 
of Eq. \ref{cavfss} in $d=1$ for two temperatures, one below and one above
$T_{gap}$.
In both cases, the agreement with the droplet scaling ansatz
is very good, confirming the zero-temperature fixed point picture.
The corresponding  finite-size scaling analysis 
for the disorder averaged correlation
in $d=3$ is shown on Fig. \ref{fig3dcavt0.1}.

\subsection{ Typical  correlation $\overline{ \ln C_L( r )}$  }

\begin{figure}[htbp]
\includegraphics[height=6cm]{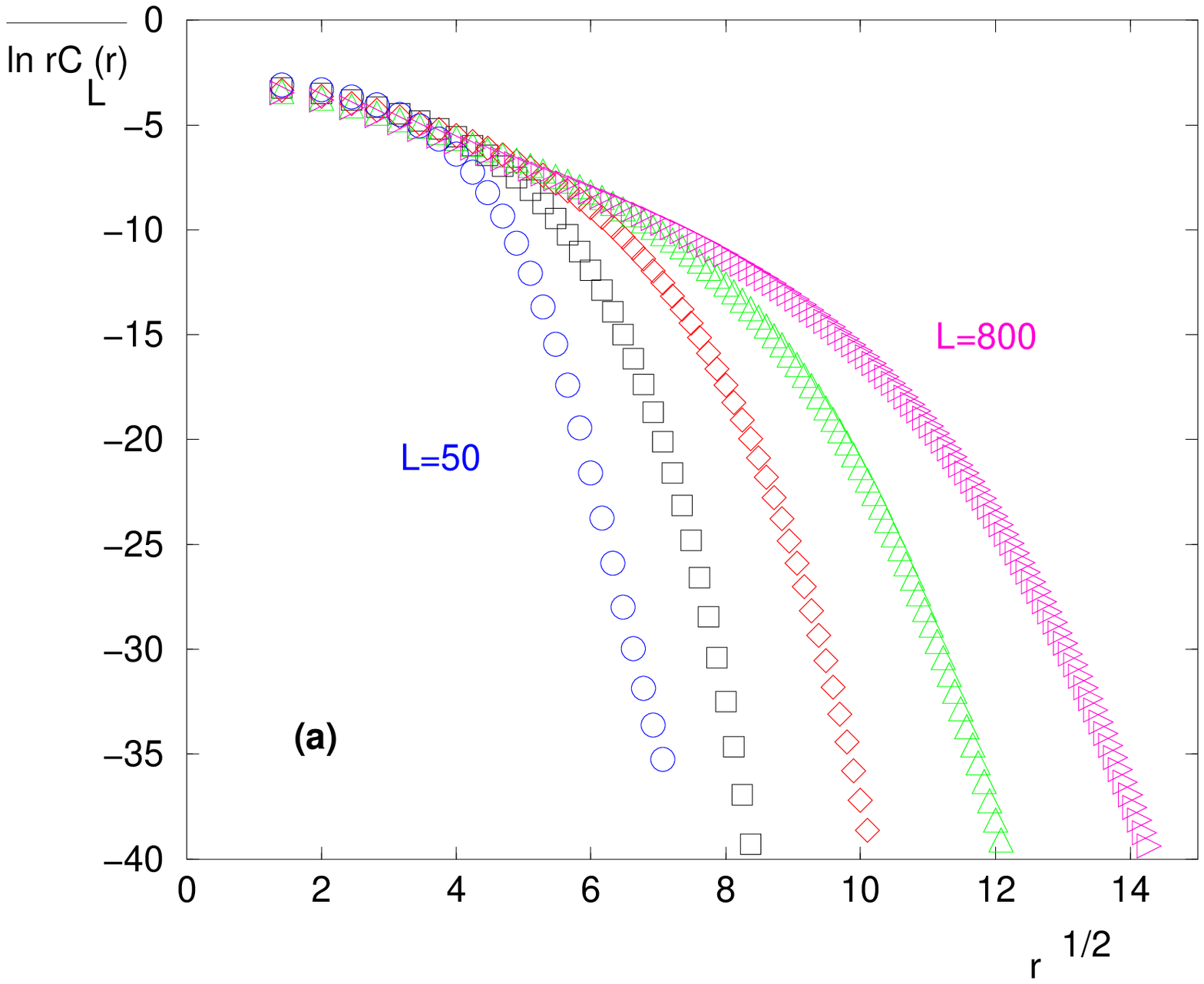}
\hspace{1cm}
\includegraphics[height=6cm]{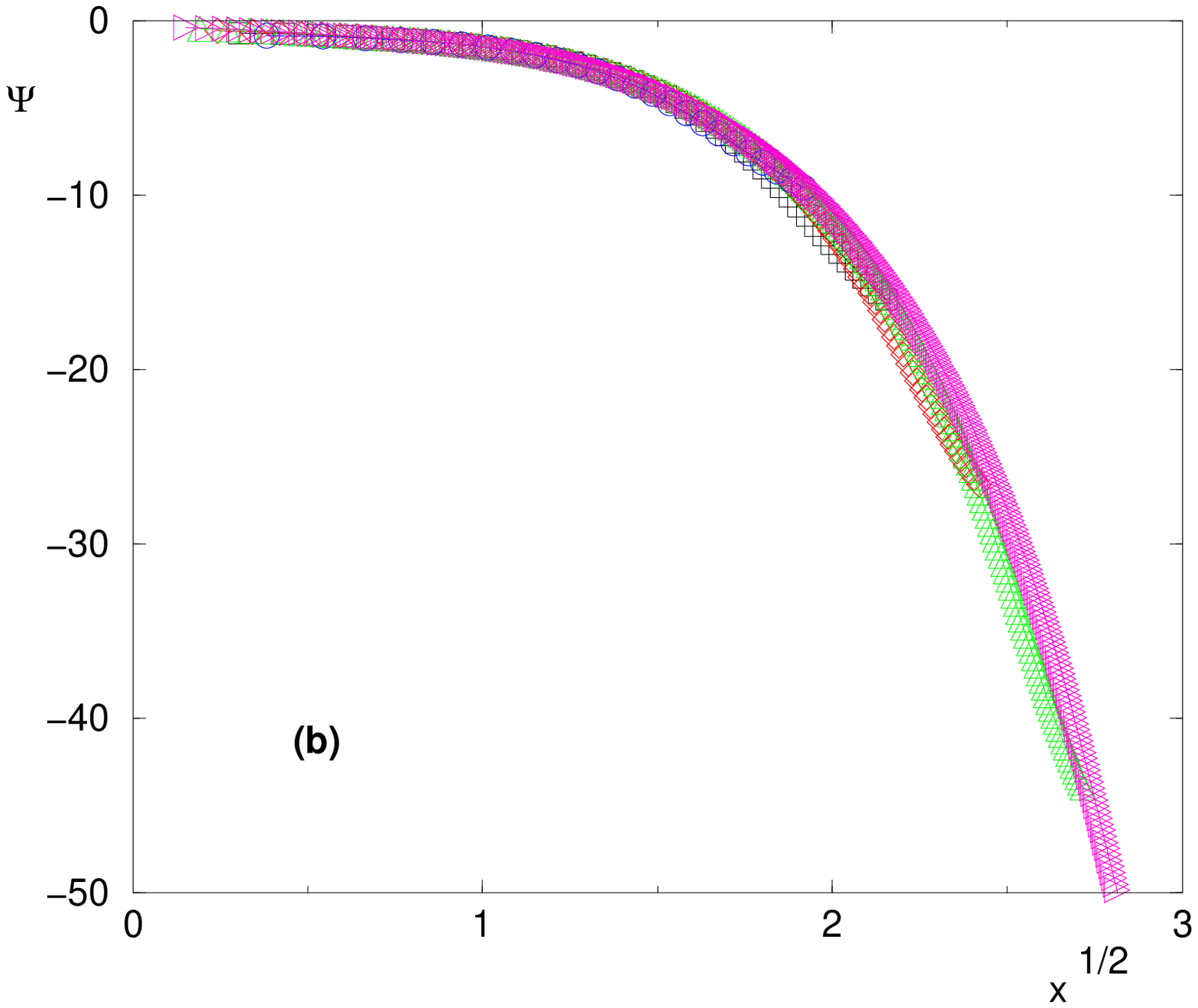}
\caption{(Color online) 
Typical correlation $\overline{ \ln C_L( r )}$ in $d=1$ for $T=1$
and sizes  $L=50 (\bigcirc),100 (\square),200
(\Diamond),400(\triangle),800 (\rhd)$ :
(a) $(\overline{ \ln C_L( r )}+\ln (r))$ as a function of $r^{1/2}$
 (see Eq. \ref{ctypLinfty} )
(b) finite-size scaling of the same data :
$\Psi=(\overline{ \ln C_L( r )}+\ln (r))/L^{1/3}$
 as a function of $x^{1/2}=(r/L^{2/3})^{1/2}$ (see Eq. \ref{ctypfss}). }
\label{fig1dclogav}
\end{figure}

\begin{figure}[htbp]
\includegraphics[height=6cm]{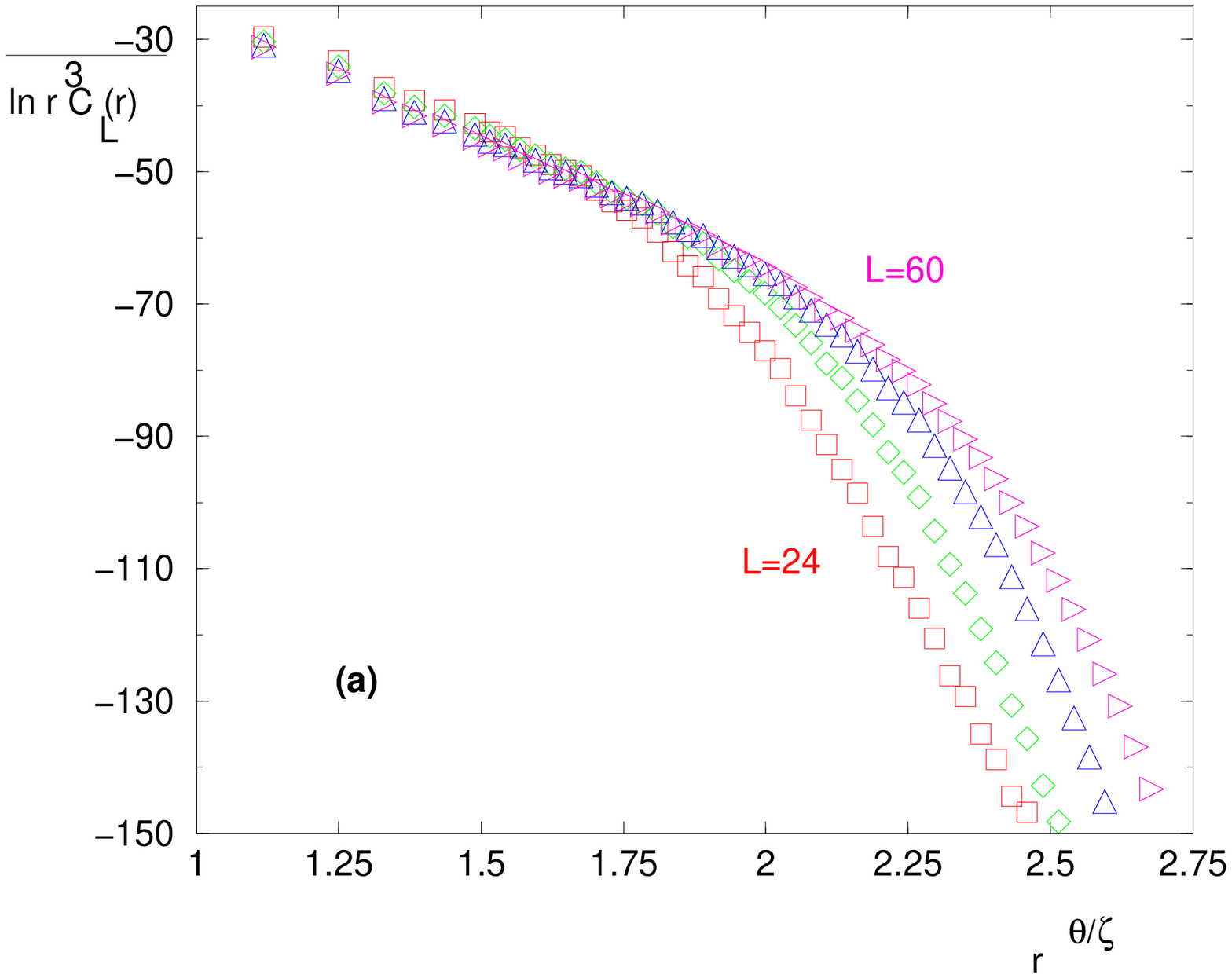}
\hspace{1cm}
\includegraphics[height=6cm]{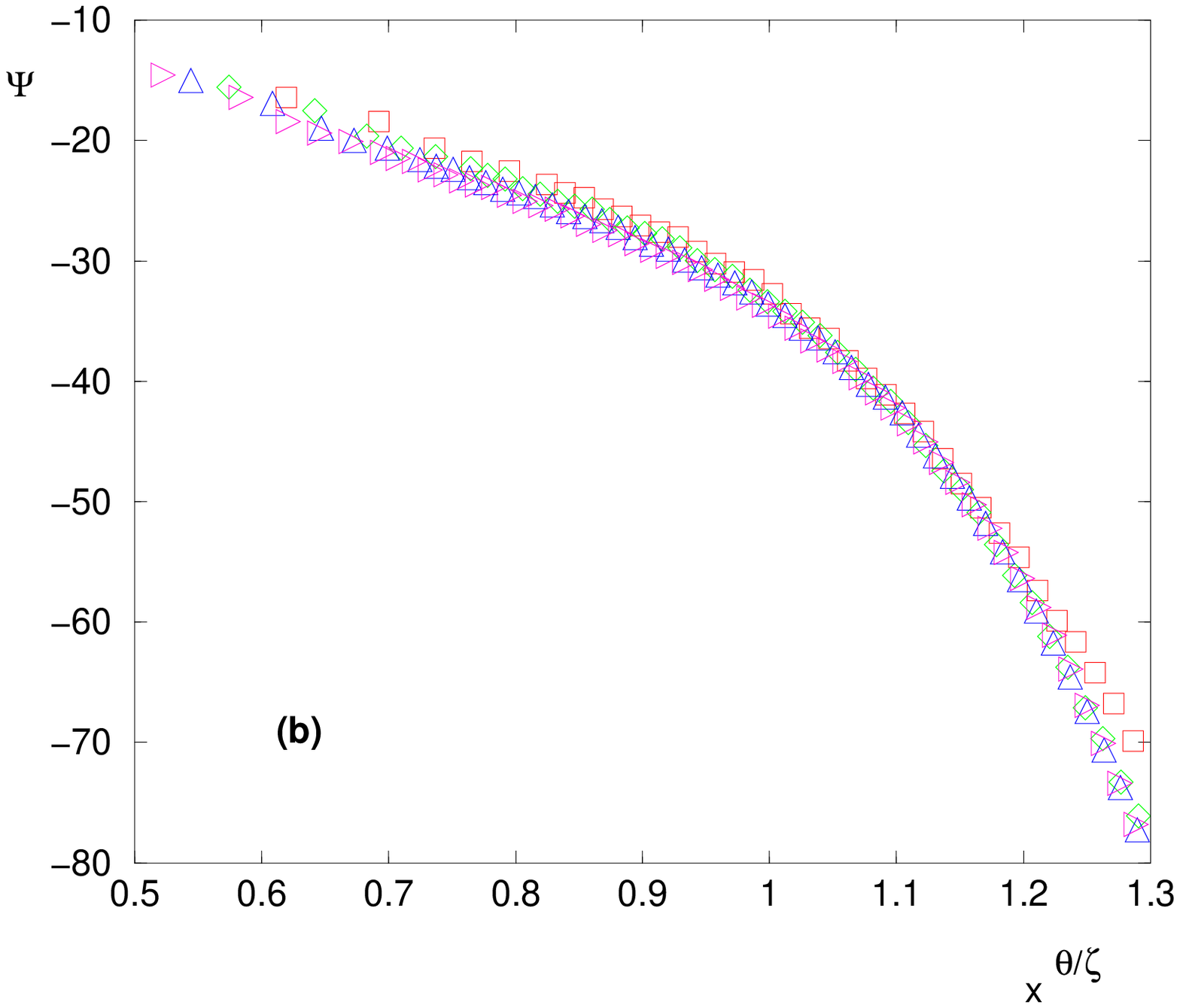}
\caption{(Color online)
 Typical correlation $\overline{ \ln C_L( r )}$ in $d=3$ for $T=0.1$
and sizes $L=24 (\square), 36 (\Diamond),48(\triangle) ,60(\rhd) $
(a) $(\overline{ \ln C_L( r )}+d \ln (r))$ 
as a function of $r^{\theta/\zeta=0.314}$
(see Eq. \ref{ctypLinfty}) 
(b) finite-size scaling of the same data :
$\Psi=(\overline{ \ln C_L( r )}+d \ln(r))/L^{\theta}$
 as a function of $x^{\theta/\zeta}=(r/L^{\zeta})^{\theta/\zeta}$
 (see Eq. \ref{ctypfss})  .}
\label{fig3dclogavt0.1}
\end{figure}

The typical correlation of Eq. \ref{ctypLinfty}
are shown on Fig. \ref{fig1dclogav} a for $d=1$
and on Fig. \ref{fig3dclogavt0.1} a for $d=3$ respectively :
the collapse for small $r$ is satisfactory.
To take into account the $L$ dependent cut-off for large $r$, 
we have tried the following finite-size scaling form
\begin{equation}
(\overline{ \ln C_L( r ) } + d \ln r) \sim  L^{\theta} 
\Psi \left( x= \frac{r}{L^{\zeta} } \right)
\label{ctypfss}
\end{equation}
where the scaling function $\Psi(v)$ behaves as the following
power-law 
\begin{equation}
\Psi(v) \oppropto_{v \to 0} v^{\frac{\theta}{\zeta}}
\label{psiv}
\end{equation}
 to recover Eq. \ref{ctypLinfty} as $L \to \infty$.

The results for $d=1$ and $d=3$ are shown on Fig. \ref{fig1dclogav} b
and \ref{fig3dclogavt0.1} b respectively.

\subsection{ Histograms of correlation function }

\begin{figure}[htbp]
\includegraphics[height=6cm]{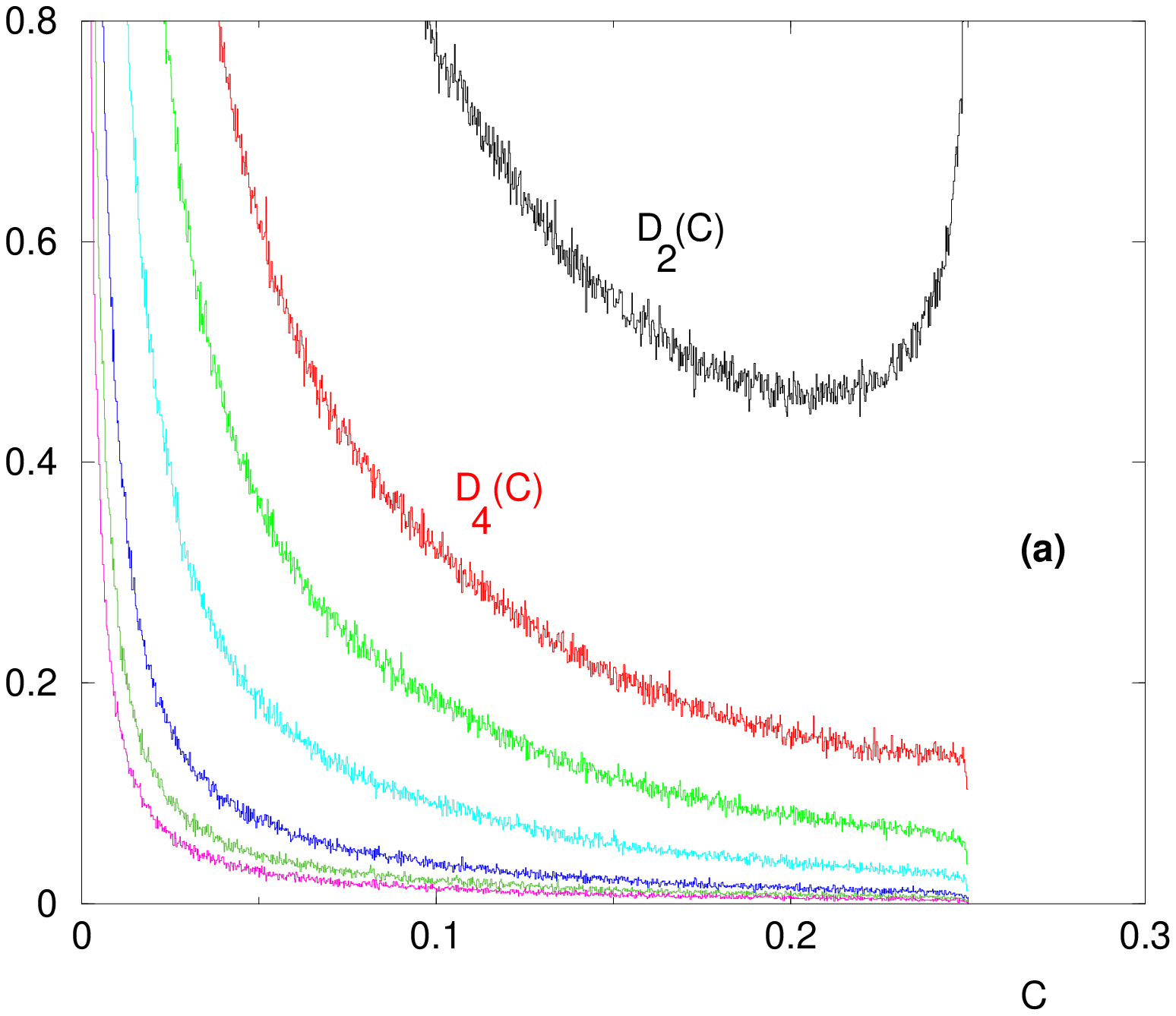}
\hspace{1cm}
\includegraphics[height=6cm]{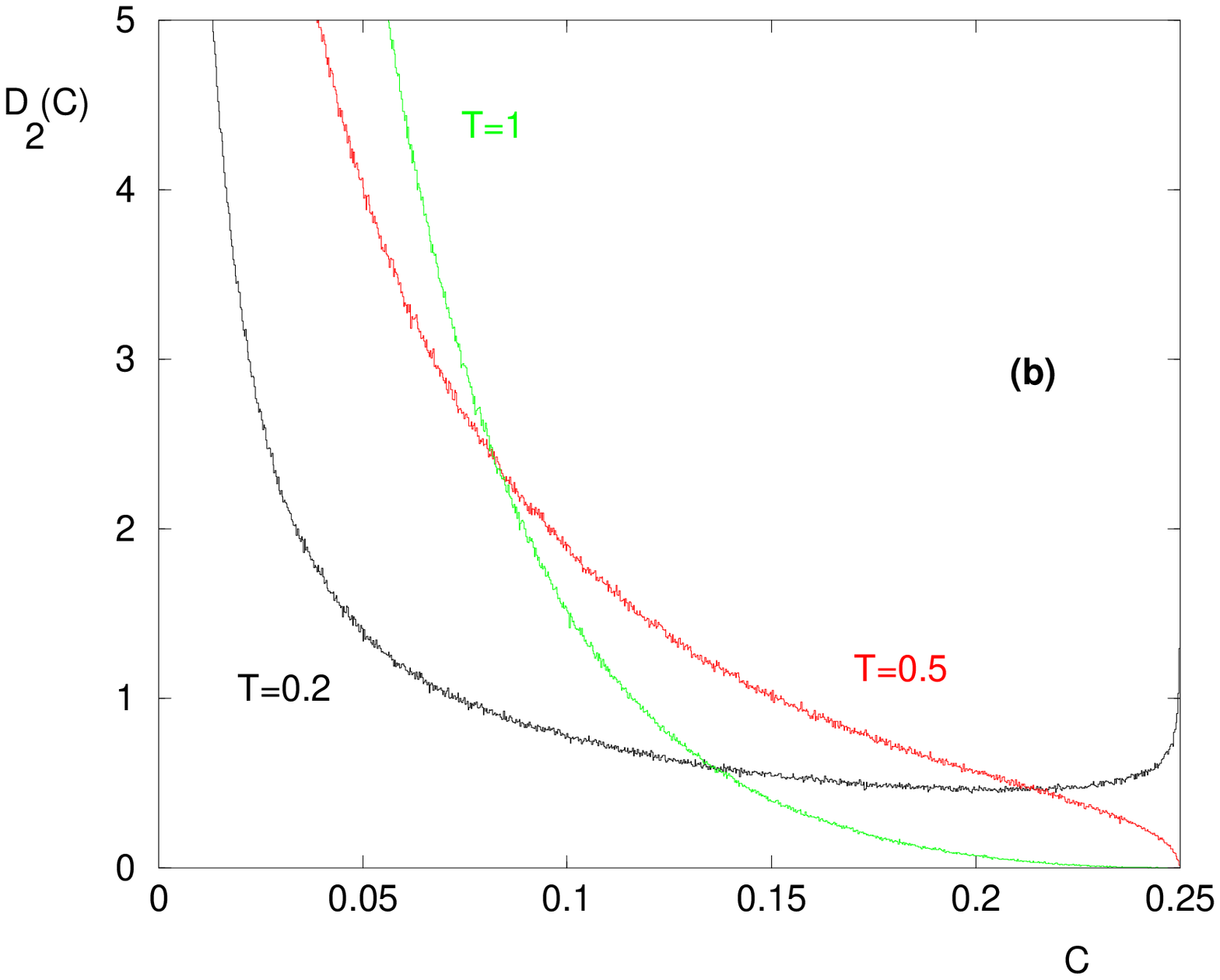}
\caption{(Color online) $d=1$ : Probability distribution $D_r(C) $
of the correlation $C(r)$ (Eq. \ref{corre}) for $L=200$
(a) for $T=0.2<T_{gap}$ the distribution  $D_r(C) $ reaches the maximal value
$C_{max}=1/4$ for any distance $r$ : here $r=2,4,6, 10, 20,30,40$
(b) Probability distribution $D_{r=2}(C) $ of the correlation of two
neighboring sites for $T=0.2$ where $\mu(T)<1$,
$T=0.5$ where $1<\mu(T)<2$ and $T= 1.>T_{gap}$ where the histogram
does not reach the maximal possible value $C_{max}=1/4$.
.}
\label{fig1dhistocorre}
\end{figure}

Since typical and disorder averaged correlations are very different,
we have also studied in $d=1$ the
probability distribution $D_r(C) $
of the correlation $C(r)$ (Eq. \ref{corre}) 
between the preferred position and a site at transverse distance $r$.
For $T<T_{gap}$, the distribution  $D_r(C) $ reaches the maximal
possible value
$C_{max}=1/4$ for any distance $r$ as shown on
Fig. \ref{fig1dhistocorre} a.
This maximal value $C_{max} \sim 1/4$ corresponds to the case
where the maximal weight and second maximal weight are 
both of order $w \sim 1/2$ and are at distance $r$.
For large $r \sim L^{2/3}$, this corresponds to the droplet
excitations.

For $T>T_{gap}$, the distribution  $D_r(C) $ 
does not reach the maximal possible value $C_{max}=1/4$ anymore,
as shown on Fig.  \ref{fig1dhistocorre} b.

\section{Conclusion}

\label{conclusion}

We have studied the weight statistics in
the low-temperature $T<T_c$ disorder-dominated phase
of the directed polymer in a random potentiel in dimension
$1+1$ (where $T_c=\infty$) and $1+3$ (where $T_c<\infty$). 
In particular, we have found a temperature $T_{gap}<T_c$ 
with the following properties.
For $T<T_{gap}$, the histograms of weight observables
and of spatial correlation display characteristic Derrida-Flyvbjerg
singularities.
 In particular, there exists a
temperature-dependent exponent $\mu(T)$ that governs the main singularities
of $P_1(w) \sim (1-w)^{\mu(T)-1}$, $\Pi(Y_2) \sim (1-Y_2)^{\mu(T)-1}$
and $G(s) \sim s^{\mu(T)-1}$.
as well as the power-law decay of the  moments $ \overline{Y_k(i)}
\sim 1/k^{\mu(T)}$.  The exponent $\mu(T)$ grows from the value
$\mu(T=0)=0$ up to $\mu(T_{gap}) \sim 2$.
For $T_{gap}<T<T_c$, the
distribution $P_1(w)$ vanishes at some value $w_0(T)<1$, and
accordingly the moments $\overline{Y_k(i)}$ decay exponentially as
$(w_0(T))^k$ in $k$. 
Finally, our numerical results concerning typical and averaged
correlations are in full agreement with the droplet scaling theory
both below and above $T_{gap}$.

Together with our previous study on the freezing transition
of random RNA secondary structures \cite{rnapoids},
this shows that the weight statistics is an efficient tool
to characterize to which extent local degrees of freedom
are frozen. Moreover, the position of $T_{gap}$ with
respect to $T_c$ gives a better understanding of the transition.
In the RNA case where $T_c<T_{gap}$, the interpretation
is that there exists frozen pairs in the high-temperature phase,
but only of finite size \cite{rnapoids}.
Here, in the directed polymer case where $T_{gap}<T_c$,
there exists frozen monomers only below $T_{gap}$,
whereas for $T_{gap}<T<T_c$, the localization occurs on a tube
of finite extent $\xi_{\perp}(T)$.

\end{document}